\documentclass[iop]{emulateapj}
\shorttitle{EVOLUTION OF THE SFR OF GALAXIES AT 0.0 $\leq z \leq$ 1.2}
\shortauthors{RUJOPAKARN ET AL.}

\newcommand{\LMIPS}{\mbox{$L(24~\micron)$}}
\newcommand{\Lsun}{\mbox{$L_\odot$}}
\newcommand{\LTIR}{\mbox{$L({\rm TIR})$}} 

\newcommand{\SFRD}{\mbox{$\dot{\rho}_*$}} 
\newcommand{\Bootes}{Bo\"{o}tes} 

\slugcomment{Accepted for publication in the Astrophysical Journal, June 8th, 2010}
\begin{document}
\title{THE EVOLUTION OF THE STAR FORMATION RATE OF GALAXIES AT $0.0 \leq z \leq 1.2$}

\author{WIPHU RUJOPAKARN\altaffilmark{1},
DANIEL J. EISENSTEIN\altaffilmark{1},
GEORGE H. RIEKE\altaffilmark{1},
CASEY PAPOVICH\altaffilmark{2},
RICHARD J. COOL\altaffilmark{3,11},
JOHN MOUSTAKAS\altaffilmark{4},
BUELL T. JANNUZI\altaffilmark{5},
CHRISTOPHER S. KOCHANEK\altaffilmark{6},
MARCIA J. RIEKE\altaffilmark{1},
ARJUN DEY\altaffilmark{5},
PETER EISENHARDT\altaffilmark{7},
STEVE S. MURRAY\altaffilmark{8},
MICHAEL J. I. BROWN\altaffilmark{9},
EMERIC LE FLOC'H\altaffilmark{10}}

\altaffiltext{1}{Steward Observatory, The University of Arizona,
  Tucson, AZ 85721; wiphu@as.arizona.edu}
\altaffiltext{2}{Department of Physics and Astronomy, Texas A\&M
  University, College Station, TX 77843-4242.} 
\altaffiltext{3}{Department of Astrophysical Sciences, 4 Ivy Lane,
  Peyton Hall, Princeton University, Princeton, NJ 08544}
\altaffiltext{4}{Center for Astrophysics and Space Sciences,
  University of California, San Diego, 9500 Gilman Drive, La Jolla,
  CA, 92093}
\altaffiltext{5}{National Optical Astronomy Observatory, Tucson, AZ
  85726.} 
\altaffiltext{6}{Department of Astronomy and the Center for Cosmology and
  Astroparticle Physics, The Ohio State University, 140 W. 18th Ave.,
  Columbus, OH 43210.} 
\altaffiltext{7}{Jet Propulsion Laboratory, Mail Stop 169-506,
  California Institute of Technology, Pasadena, CA 91109.} 
\altaffiltext{8}{Harvard/Smithsonian Center for Astrophysics, Mail
  Stop 67, 60 Garden Street, Cambridge, MA 02138.} 
\altaffiltext{9}{School of Physics, Monash University, Clayton,
  Victoria 3800, Australia.} 
\altaffiltext{10}{Institute for Astronomy, University of Hawaii,
  Honolulu, HI 96822.}
\altaffiltext{11}{{\it Hubble} Fellow}

\begin{abstract}
We present the 24 $\micron$ rest-frame luminosity function (LF) of
star-forming galaxies in the redshift range $0.0 \leq z \leq 0.6$
constructed from 4047 spectroscopic redshifts from the AGN and Galaxy 
Evolution Survey of 24 $\micron$-selected sources in the
\Bootes\ field of the NOAO Deep Wide-Field Survey. This sample provides
the best available combination of large area (9 deg$^2$),
depth, and statistically complete spectroscopic observations, allowing
us to probe the evolution of the 24 $\micron$ LF of galaxies at low
and intermediate redshifts while minimizing the 
effects of cosmic variance. In order to use the observed 24 $\micron$
luminosity as a tracer for star formation, active galactic nuclei (AGNs) that
could contribute significantly at 24 $\micron$ are identified and excluded
from our star-forming galaxy sample based on their mid-IR spectral
energy distributions or the detection of X-ray emission. Optical
emission line diagnostics are considered for AGN identification, but
we find that 24 $\micron$ emission from optically-selected AGN is 
usually from star-forming activity and therefore should not be
excluded. The evolution of the 24 $\micron$ LF of star-forming galaxies
for redshifts of $z \leq 0.65$ is consistent with a pure luminosity
evolution where the characteristic 24 $\micron$ luminosity evolves as
$(1+z)^{3.8 \pm 0.3}$. We extend our evolutionary study to
encompass $0.0 \leq z \leq 1.2$ by combining our data with that of the
Far-Infrared Deep Extragalactic Legacy Survey. Over this
entire redshift range the evolution of the characteristic 24 $\micron$
luminosity is described by a slightly shallower power law of
$(1+z)^{3.4 \pm 0.2}$. We find a local star formation rate density of
$(1.09 \pm 0.21) \times 10^{-2}$ $M_{\odot}{\rm yr}^{-1}{\rm
  Mpc}^{-3}$, and that it evolves as $(1+z)^{3.5 \pm 0.2}$ over $0.0
\leq z \leq 1.2$. These estimates are in good agreement with the rates
using optical and UV fluxes corrected for the effects of intrinsic
extinction in the observed sources. This agreement confirms that star
formation at $z \lesssim 1.2$ is robustly traced by 24 $\micron$
observations and that it largely occurs in obscured regions of
galaxies.
\end{abstract}
\keywords{cosmology: observations --- galaxies: evolution --- galaxies: 
high-redshift --- infrared: galaxies}

\section{INTRODUCTION}
We have learned in the past 15 years that the star formation rate
(SFR) of the universe peaked at a redshift of $z \geq 1$ and then
rapidly declined by an order of magnitude to the present day
\citep[e.g.,][]{Lilly96, Mad96, CharyElbaz01, Hopkins04, LeFloch05,
  PPG05, HopkinsBeacom06} . The exact cause of the decline of the SFR
remains unknown. However, since the period from $z \sim 0.6$ to the
present encompasses nearly half of cosmic time coinciding with the SFR
decline, this redshift range is a key regime for constraining the
mechanisms that may be responsible. 

The evolution of the SFR has been measured extensively at mid-IR (MIR;
8$-$30 $\micron$), far-IR (FIR), ultraviolet (UV), and optical
wavelengths. Among these wavelengths, MIR and UV are ideal tracers for
dust-obscured and unobscured galaxies, respectively. The MIR emission
of galaxies originates from the reprocessed UV radiation of
dust-embedded, highly obscured star-forming regions, and hence is
sensitive to radiation from star formation that can not be seen
directly at optical wavelengths \citep[e.g.,][]{Kennicutt98, Buat99}. The
stellar UV radiation field is largely created by hot OB stars ($M_* >
10~M_{\odot}$), which have very short lifetimes ($<20$ Myr) and hence
trace current star formation. The dust obscuration is generally
substantial for star formation in galaxies at and above the
characteristic luminosity L$_*$ \citep{Schmitt06, Buat07}. Such
galaxies comprise the majority of galaxies studied beyond the low
redshift volume, so the MIR offers a relatively complete probe of SFR
at intermediate and high redshifts. Among the IR wavelengths, the 24
$\micron$-band is particularly suitable to study SFR evolution because
it avoids most of the strong emission features from the stochastically
heated aromatic hydrocarbons (hereafter PAH) at wavelengths $3.3 <
\lambda < 13~\micron$ for redshifts $z \lesssim 0.5$ while still
avoiding the contamination from the Galactic IR cirrus at longer
wavelengths \citep{Shupe98}. 

Measurements of IR number counts and luminosity functions (LFs) of
galaxies demonstrate the rapid evolution of the SFR. Early IR studies
utilized the {\it IRAS} Bright Galaxy Sample \citep{Soifer89} and the
{\it IRAS} calibration field \citep{HackingHouck87} and focused on 
accounting for the cosmic infrared background \citep[CIB;][]{Elbaz99}
with discreet MIR sources \citep{Hacking87, Saunders90,
  SoiferNeu91}. Studies of galaxy number counts with {\it ISO} and {\it
  Spitzer} have shown that IR galaxies evolve very strongly
\citep{Elbaz99, Frances01, Dole04, Marleau04, Papovich04}. As
redshifts of IR samples became available, the LF could be 
constructed in the nearby volume \citep{RiekeLebof86, Saunders90,
  Yahil91, RowRob97, Shupe98} and subsequently extended to
intermediate and high redshifts ($z > 3$) with large samples of 
photometric redshifts \citep{Pozzi04, LeFloch05, PPG05, Caputi07,
  Marleau07} to study the SFR evolution over cosmic time. However,
most of these studies were either in the local regime ($z<0.1$) or at
intermediate to high redshifts ($z>0.5$). The redshift range $0.1 < z
< 0.6$ includes more than 4 Gyr of galaxy evolution and needs to be
characterized in detail to understand this process.

Observations of galaxy evolution at low and intermediate redshifts
face two challenges. First, a large area, deep survey is required to 
access a large volume at lower redshifts, which is critical to
minimize cosmic variance. For example, a deep survey at redshift $z
\sim 0.5$ would need five times the solid angle coverage of one at $z
\sim 1$ to probe a comparable comoving volume. Second, while
photo-$z$'s are generally adequate for distance determination at high
redshift, accurate spectroscopic redshifts are critical at low and
intermediate redshifts. To illustrate this issue, consider typical
photometric redshift uncertainties of $\Delta z/(1+z) = 0.05$. The
corresponding fractional uncertainty in $z$ would be a tolerable 10\%
at $z \sim 1$, while at $z \sim 0.25$ and below, the 
fractional uncertainty would exceed 25\%. Also, photometric redshifts
are only accurate for galaxies whose spectral energy distributions
(SEDs) are well matched 
by either SED templates or, in the case of empirical photometric redshifts, 
galaxies in a spectroscopic training set. This may not be the case for
heavily obscured galaxies such as luminous IR galaxies (LIRGs). In
addition, the accuracy of some photometric redshift samples in the
literature has not been verified with spectroscopy, and unknown
systematic errors may be present. 

In this work we construct the rest-frame 24 $\micron$ LFs of
star-forming galaxies, measure their evolution, and estimate the SFR
for $0.0 \leq z \leq 0.6$ from the LF. In addition to constraining the
LF and SFR evolution accurately at low and intermediate redshifts, we
combine our results with those from the Far-IR Deep Extragalactic
Legacy Survey \citep[FIDEL,][]{Magnelli09} to extend the constraints
to $z \sim 1.2$. We organize this paper as follows. In
Section \ref{sec:data} we discuss our  {\it Spitzer} 24 $\micron$ sample, 
the spectroscopic redshifts from AGN and Galaxies Evolution Survey
(AGES), the optical catalog from NDWFS, 
and address their completeness. In Section \ref{sec:IR_LF} we describe the 
construction of the LF, starting with the derivations of galaxy
luminosities (Section \ref{sec:IR_LF_LI_L24_Estimation}), identification
and exclusion of AGN from star-forming galaxies
(Section \ref{sec:IR_LF_AGN}), estimators for the LF 
(Section \ref{sec:IR_LF_Method}), parameterization of the evolution of the LF
with redshift (Section \ref{sec:IR_LF_evolparam}), and the estimation of the SFR 
from the LF (Section \ref{sec:IR_LF_SFR}). We then present the results in
Section \ref{sec:results} for the local LF (LLF; Section
\ref{sec:results_LLF}), the 
LF evolution at low and intermediate redshifts
(Section \ref{sec:results_LF_evol}), the LF evolution constraint out to $z
\sim 1.2$ (Section \ref{sec:results_LF_evol_to_FIDEL}), as well as the
SFR evolution of star-forming galaxies (Section
\ref{sec:results_SFRD}). We summarize our results in 
Section \ref{sec:conclusions}. We assume a $\Lambda$CDM cosmology with
$\Omega_m = 0.3$, $\Omega_{\Lambda} = 0.7$, and $H_0 =
70$~km~s$^{-1}$Mpc$^{-1}$ throughout this paper.  

\section{THE DATA}\label{sec:data}

We construct the 24 $\micron$ LF of galaxies in the 9 sq. degree
\Bootes\ field of the NOAO Deep Wide-Field Survey \citep[NDWFS Data
  Release 3; $\alpha=14^h32^m$, $\delta=+34\arcdeg16\arcmin$,
  J2000.0;][]{Jannuz99}. The \Bootes\ field has optical
imaging reaching the 5-$\sigma$ flux limit of  26.6, 26.0, 26.0, and
21.4 in $B_W$, $R$, $I$, and $K$ (Vega magnitudes),
respectively. Details about the NDWFS can be found in \citet{Jannuz99}
and at the survey's
Web site\footnote[1]{http://www.noao.edu/noao/noaodeep}. 

In addition to the deep optical imaging, the \Bootes\ field has
extensive panchromatic coverage from other surveys, including an X-ray
survey with the {\it Chandra} Space Telescope
\citep[XBo\"{o}tes;][]{Murray05} comprising of 5 ks ACIS-I exposures,
near-UV deep 30 ks imaging from {\it GALEX} \citep{Martin03}, the IRAC
Shallow Survey in the 3.6, 4.5, 5.8, and 8.0 $\micron$ bands reaching
magnitudes of 19.1, 18.3, 15.9, and 15.2 mag (Vega mag), respectively,
\citep{Eisenhardt04}, the deeper {\it Spitzer} Deep, Wide-Field Survey
\citep[SDWFS,][]{Ashby09}, and MIPS MIR photometry at 24 $\micron$ down
to 0.27 mJy \citep{Papovich04, Houck05}. These imaging
data are complemented by the set of optical spectroscopic redshifts
from the AGN and Galaxy Evolution Survey (AGES; Kochanek et al., in
prep.), which is a statistically complete redshift survey of the field
at low and intermediate redshifts ($0 \leq z \leq 0.8$) (C. S. Kochanek et
al., in preparation). 

\subsection{{\it Spitzer} 24 $\micron$-Observations, Reduction, and the
  Final Catalogue}\label{sec:data_MIPS}

The 24 $\micron$ photometry in the \Bootes\ field was obtained by
the {\it Spitzer Space Telescope} wide-area survey with the Multiband
Imaging Photometer for {\it Spitzer} \citep[MIPS;][]{Rieke04} and
processed with the MIPS GTO Data Analysis Tool
\citep[DAT;][]{Gordon05}. The mean exposure time per pixel across the
\Bootes\ field is 87 s. The pixel size of the final mosaic is
$1\farcs25$ with a FWHM of $5\farcs7$.

Source extraction from the final 24 $\micron$ mosaic was done using the 
techniques described in \citet{Papovich04}. Briefly, we weight the
image by the typical integration time of each region to obtain uniform
noise across the mosaicked image. We filtered the image with a
Gaussian kernel similar to our $24~\micron$ point-spread function
(PSF) to remove artifacts 
and optimize point-source detection and then used DAOPHOT
\citep{Stetson87} to detect sources. The actual photometric
measurement was carried out by fitting empirical PSF constructed from
bright sources in the image. Based on 
Monte Carlo simulations by \citet{Papovich04} the 80\% completeness
flux threshold is $F_\nu(24~\micron) = 0.27$ mJy, and there are 8,572
objects in our sample above this flux level.

The MIPS survey of the \Bootes\ field combines considerable depth
with a large contiguous area of 32,457 arcmin$^2$ (9.0 deg$^2$). While
many other fields have been surveyed to deeper $F_\nu(24~\micron)$,
such as the GOODS, ELAIS and CDF-S fields \citep{Pozzi04, LeFloch05,
  Chary07, Elbaz07} with 80\% completeness flux of 25 $\mu$Jy, 61
$\mu$Jy and 83 $\mu$Jy, respectively, they also cover significantly
smaller area (0.011 deg$^2$, 0.036 deg$^2$, and 0.407 deg$^2$,
respectively), and therefore encompass significantly smaller volumes
at low and intermediate redshifts. The Cosmic Evolution Survey
\citep[COSMOS,][]{Scoville07} has a large sample of spectroscopic
redshifts \citep{Lilly07} and deep, $\sim$80 $\mu$Jy, 24 $\micron$
imaging \citep{LeFloch09}. The 2 deg$^2$ area of the COSMOS field is
one of the best combinations of large area and depth at higher
redshifts; however, a larger angular coverage is still
required to suppress the effects of cosmic variances at low and
intermediate redshifts.  The available large-area MIR surveys, such as
the {\it Spitzer} Wide-area Infrared Extragalactic Survey (SWIRE) and the
European Large Area {\it ISO} Survey (ELAIS, 15 $\micron$), currently
lack thorough spectroscopic redshifts. 

\subsection{Optical Spectroscopy Target Selection, Observation, and
  the Final Catalogue}\label{sec:data_AGES}

The \Bootes\ field was observed spectroscopically by the AGES during
$2004$-$2006$ using the Hectospec spectrograph in a survey region
of 7.9 deg$^2$ (C. S. Kochanek et al., in preparation). Hectospec is a
fiber-fed spectrograph with 300 robotically assigned fibers covering a
1-degree diameter field of view at the $f$/5 focus of the 6.5-meter
MMT Observatory \citep{Fab05}. AGES produced $\sim$20,000 redshifts,
of which $\sim$16,500 and $\sim$3,500 are galaxies and quasars,
respectively, with redshift uncertainties less than $\sigma_z \sim 100
{~\rm km~s}^{-1}$ (as in the similar survey by \citet{Papovich06}) and
a median galaxy redshift of $z = 0.26$ (C. S. Kochanek et al., in
preparation). 

Targets for spectroscopy were principally selected on the basis of
their optical magnitudes. AGES targeted galaxies to $I_C < 20$, where
$I_C$ is the Kron-like magnitude (in Vega units) in the Kron-Cousins
I-band of the NDWFS \citep[SExtractor's {\tt
    MAG\_AUTO};][]{Bertin96}. For galaxies brighter than $I_C=18.5$, 
there was full sampling.  For fainter galaxies with $18.5 < I_C < 20$,
AGES used random sparse 
sampling of 1 in 5 galaxies, with rates that depended on a variety of
other factors, so as to put emphasis on more unusual objects. Additional
targets were added to the spectroscopic sample if they were brighter
than certain flux limits in the X-ray, infrared, other optical bands,
and radio. For example, all galaxies with $F_\nu(24~\micron)$ above
0.5 mJy were targeted, as well as 30\% of those down to 0.3 mJy. For
each object, the a priori sampling rate is known and so the object can
be exactly weighted to form an unbiased sample of $I_C<20$ galaxies.
AGES observed the Bo\"otes field with 15 complete tilings, so the
completeness for the target galaxies is very high. Our final sample
has 4.3\% incompleteness due to difficulties in assigning 
fibers to targets and 2.1\% incompleteness due to failing to get a
successful redshift from an assigned fiber. We also estimate that the
parent photometry catalog is missing 4\% of the objects in the range
$15 < I_C < 20$, half of which is due to interference by nearby bright
stars, the rest likely due to blending and corruption of the parent
photometry.  We model these mild incompletenesses, including factors
for areal target density and surface brightness, and correct the final
weights of objects in the redshift catalog (R. J. Cool et al., in
prep.).  As the total incompleteness correction is only 10\%, it is
very likely that the systematic error in the correction is far less
than the Poisson and cosmic variance in our results.

Each Hectospec fiber is $1\farcs5$ in diameter and produces a spectrum 
covering the wavelength range of 3500~\AA{} $\leq \lambda \leq$ 9200
\AA{} with 6 \AA{} FWHM resolution ($R \sim$ 1,000). Typical
signal-to-noise ratios of the AGES spectra are $2-14$ per pixel with a
median of 5 per pixel. We fit redshifts to AGES with two automated
pipelines, one based on the SDSS template-fitting code, the other on
cross-correlations. The pipelines make use of spectral features
across the spectral range (absorption lines, emission lines, and
spectral breaks due to metal-line absorptions) and hence result in
accurate redshifts regardless of whether a particular spectral line is
compromised (e.g. by overlap with sky lines). The two pipelines
gave broadly consistent 
results. In addition, we visually examined every fit. Any objects
with visually questionable results were tagged for reobservation,
since AGES covered the full field in six sequential passes.  The
result is a spectroscopic catalog with high completeness and high
reliability.  The spectroscopic success rate, after fiber assignment,
is 98\%, and we expect the redshift error rate on claimed successes to
be well below 1\%.  Any systematic biases due to errors in the
completeness corrections or the redshift catalog are likely to be
considerably smaller than the statistical errors of our LF.

In the redshift range of interest, $0.05 \leq z \leq 0.65$, we
have redshifts for 4,047 objects with $F_\nu(24~\micron)$ above the
80\% completeness limit of 0.27 mJy.

\section{INFRARED LUMINOSITY FUNCTION}\label{sec:IR_LF}
Here we describe the construction of the 24 $\micron$ LF. We start by
estimating the 24 $\micron$ and optical (I-band) rest-frame
luminosities, then identify and exclude AGN in our sample, and finally
construct the LF using the classical 1/$V_{\rm max}$ \citep{Schmidt68,
  Huchra73} and the maximum likelihood estimator \citep[MLE;
  e.g.,][]{STY79, Marshall83, Yahil91}. Later in this section, we  
describe the parameterization of LF evolution, which will be used to
study the evolution of the SFR.

\subsection{24 $\micron$ and I-band Luminosity
  Estimation}\label{sec:IR_LF_LI_L24_Estimation} 
In addition to the obvious need for 24 $\micron$ luminosities, \LMIPS,
to construct the LF, we have to compute the I-band luminosity for each
galaxy. AGES spectroscopy was targeted based on an I-band,
flux-limited parent catalog. Hence, an object can only be included
in the final LF sample if it is detected above the flux limits at both
24 $\micron$ and I-band. Therefore, the luminosities in both bands are
needed for estimating the maximum redshift, z$_{max}$, that an object
would still be detected in both bands to the construct the $1/V_{max}$
LF. \LMIPS\ is estimated by the formalism given by
\citet{Rieke09}. For the I-band, we derive the I-band k-correction
using the {\tt kcorrect} code \citep{Blanton07} and then use the 
k-correction to estimate the I-band luminosity.

The I-band k-corrections were derived using optical photometry from
NDWFS and redshifts from AGES. A fraction of our I-band objects are
affected by scattered light from bright stars. Specifically, some of
the NDWFS I-band Kron-like magnitudes (SExtractor's {\tt MAG\_AUTO}),
$I_C$, are overly bright due to contamination from
low-surface-brightness halos of nearby bright stars. This problem can
be corrected by synthesizing a new 
I-band magnitude, $I_R$, by adding a Kron-like magnitude in R-band,
$R_{C}$, to the $I-R$ color from 6$^{\prime\prime}$ apertures 
($I_R = R_C + I_{6^{\prime\prime}} - R_{6^{\prime\prime}}$). This synthesized 
$I_R$ is less sensitive to the I-band halo than $I_C$. Then we compare
$I_R$ to $I_C$ to compute a new I-band magnitude, $I_{total}$, that
uses the fainter magnitude if the two differ significantly; otherwise
we average them (R. J. Cool et al., in preparation). The $I_{total}$
computation is done in a way that the continuity of the magnitudes is
preserved. In the cases that require averaging $I_R$ and $I_C$, the
differences are on the order of a tenth of a magnitude and hence
averaging a logarithm does not introduce a significant bias. We note
that the aperture correction is insensitive to the choice of
estimation procedure and that the final LFs are little affected by 
this correction, ultimately because the I-band generally has deeper
photometry than at 24 $\micron$ and hence most of the values of 
z$_{max}$ will be limited by the 24 $\micron$ photometry. Apart from
the $I_{total}$ for I-band photometry, we use the
$4^{\prime\prime}$-diameter aperture 
magnitude from NDWFS for the $B_W$ and $R$ bands to perform
k-correction estimation. We adopt the following corrections to convert
the Vega zero points to AB: +0.02 mag, +0.19 mag, and +0.44 mag in
$B_W$, $R$, and $I$, respectively. The Galactic extinction for our
field is $A_I = 0.02$ mag \citep{SFD98}.

The estimation of \LMIPS\ requires an assumption for the spectral
energy distribution (SED) of a galaxy in the IR regime, which has been
modeled by many authors \citep[e.g.,][]{Dale01, CharyElbaz01,
  DaleHelou02, Lagache03, Lagache04, JDSmith07,Dale07}. More recently,
\citet{Rieke09} constructed a library of SEDs by combining stellar
photospheric spectra from 0.4$-$5 $\micron$ with {\it Spitzer} IRS
spectra from 5$-$30 $\micron$ and a modified blackbody spectrum at longer
wavelengths for 11 representative local 
Luminous IR galaxies (LIRGs; $10^{11} \Lsun < \LTIR < 10^{12}
\Lsun$) and ultraluminous IR galaxies (ULIRGs; $\LTIR >  
10^{12} \Lsun$). Higher \LTIR\ galaxies contain more intense
star-forming activity and hence a larger fraction of warm dust, leading
to changes in the SED with IR luminosity.

Normally, the derivation of SFR from MIR photometry involves selecting
the SED that best matches the object's luminosity given its MIR flux
and redshift, integrating this SED to obtain \LTIR, and then using
a SFR metric to calculate the SFR. However, this procedure implies
fitting a SED with only one data point, 24 $\micron$ in our case,
because the number of objects that are detected in multiple MIPS bands
is very limited, which could lead to significant
uncertainties. \citet{Rieke09} circumvented this problem by first
assigning a SFR to each SED, which has a unique \LTIR. They then  
derive a flux in the relevant band pass (24 $\micron$) over a range of
redshifts. 

\begin{figure*}
\epsscale{0.625}
\figurenum{1}
\plotone{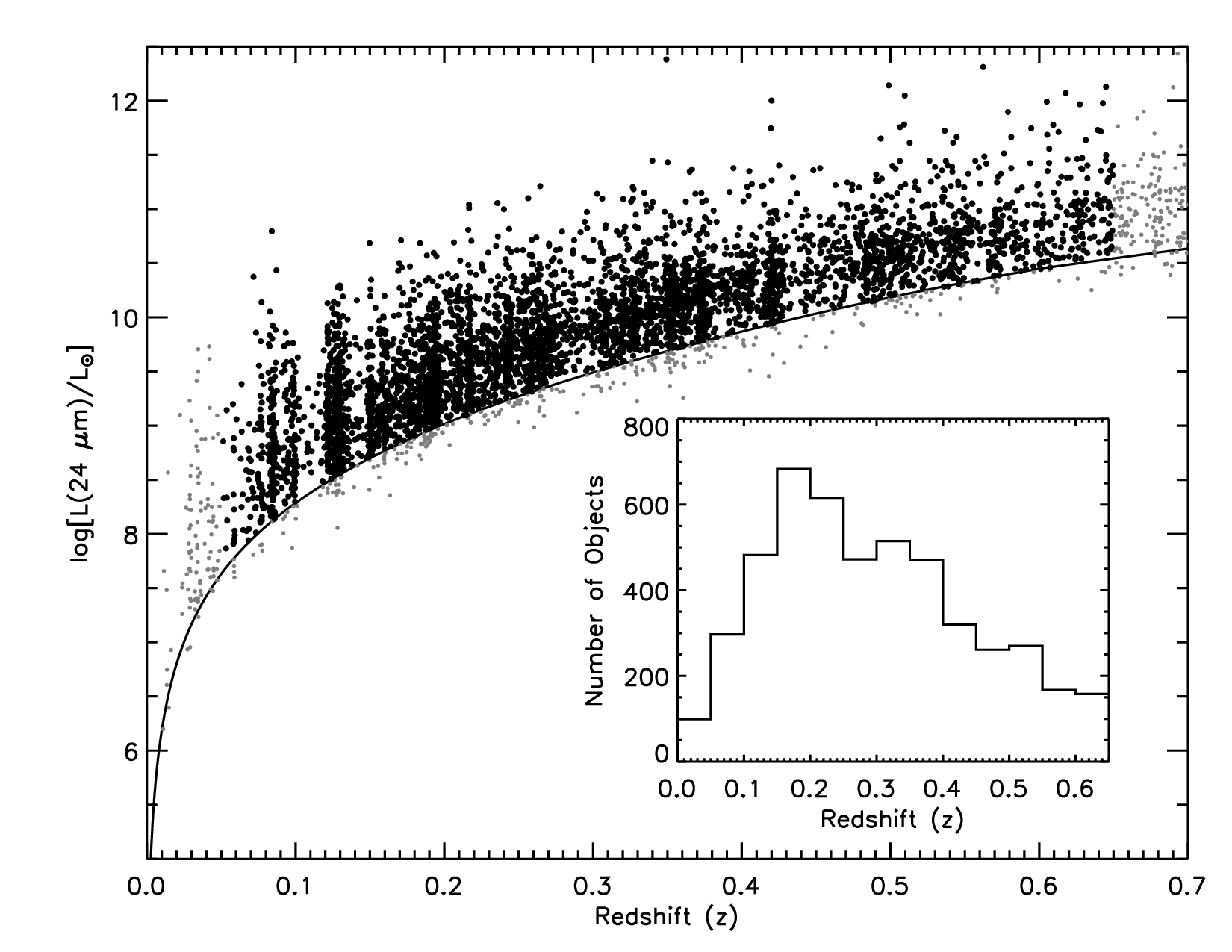}
\caption{Redshift distribution of the rest-frame 24 $\micron$
  luminosity for our sample. The gray dots show all 24 $\micron$
  galaxies with spectroscopic redshifts from AGES. The black dots are
  those above the 24 $\micron$ completeness flux limit (0.27 mJy) at
  $0.05 \leq z \leq 0.65$ that we use to construct the LF. The line
  indicates the \LMIPS\ corresponding to this
  completeness flux threshold as a function of redshift. Inset:
  the redshift distribution of the AGES objects detected at 24
  $\micron$.}
\label{Fig_Lz}

\epsscale{0.65}
\figurenum{2}
\plottwo{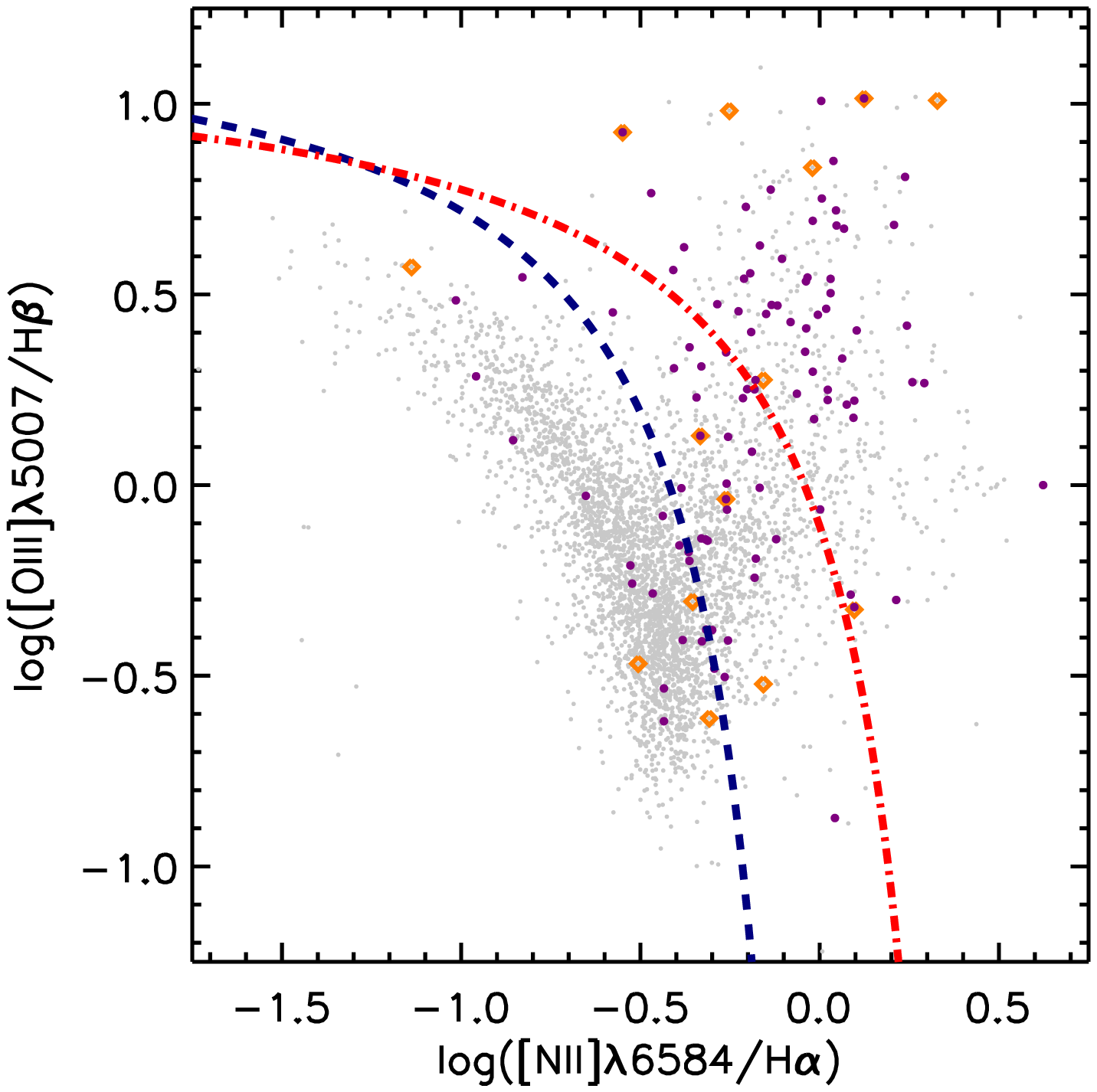}{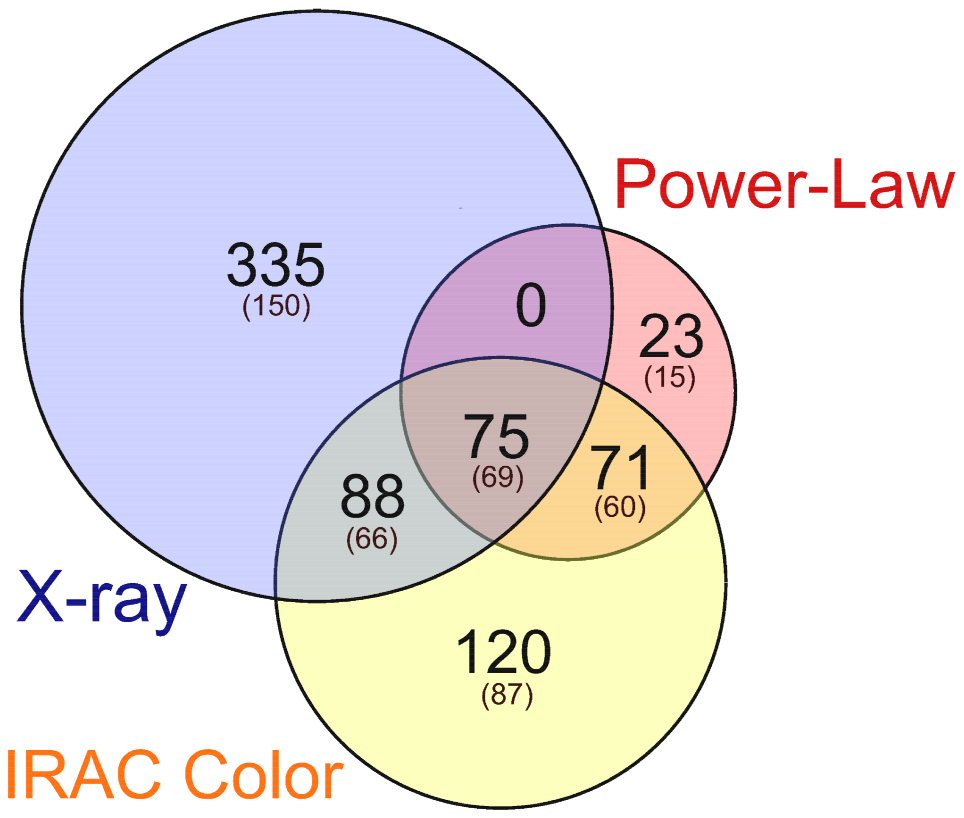}
\caption{Left: optical selection of AGN using the
  BPT diagram. Orange diamonds and magenta dots are power-law and
  X-ray selected AGNs, respectively. The Kewley and Kauffmann lines are
  shown by red dot-dashed line and blue dashed line, respectively
  \citep{Kewley01,Kauffmann03}. Many power-law and
  X-ray selected AGN are identified as star-forming galaxies by the BPT
  diagram. Note that a large fraction of power-law and X-ray selected
  AGN are not shown here because they do not have sufficiently strong
  emission lines to apply the BPT diagnostic and most of them are at
  $z > 0.35$ where H$\alpha$ line redshifts out of the Hectospec
  spectrum. Right: the Venn diagram for 712 AGNs at $0.0 \leq z \leq
  0.65$ to illustrate the overlapping of AGN identified by X-ray
  detections, power-law fitting, and the \citep{Stern05} color
  criteria. The numbers in each area and the numbers in parentheses
  are the total numbers of objects and the numbers of objects detected
  at 24 $\micron$, respectively. 288 of 712 AGNs in this Venn diagram
  are detected at 24 $\micron$, pass all quality cuts, and are in the
  main galaxy sample. They are excluded from the sample to obtain the
  star-forming galaxy sample that can be accurately used as a tracer of
  star formation.}
\label{AGN_BPT_Venn}
\end{figure*}

Given that $\LTIR/\LMIPS \sim 10$ \citep[e.g.,][]{Rieke09},
our sample with galaxies mainly in the 24 $\micron$ luminosity range
$10^{8.5} < \LMIPS < 10^{10.5}$ (Figure \ref{Fig_Lz}), is
dominated by normal star-forming galaxies ($\LTIR < 10^{11} 
\Lsun$). Figure \ref{Fig_Lz} also illustrates that LIRGs only start
contributing significantly to the MIR luminosity density at redshifts
greater than $z = 0.2$. Although ULIRGs start to appear in our sample
at $z \sim 0.4$, we will demonstrate in Section
\ref{sec:IR_LF_AGN_conclusion} that they are mostly AGN and we will
exclude them from our sample of star-forming galaxies. 

\subsection{Separating Star-Forming Galaxies and AGN Hosts}\label{sec:IR_LF_AGN}
To quantify the 24 $\micron$ emission due to star formation,
we need to minimize the contribution of AGN. There are several
methods to identify AGN in a galaxy sample. Optical emission line
diagnostics such as the BPT method \citep{BPT81}
are commonly used when optical spectra are available. MIR power-law
characteristics of AGN \citep[e.g.,][]{AH06, Donley07} and MIR color
criteria \citep[e.g.,][]{Lacy04, Stern05} can be used if multi-band MIR
photometry is available. X-ray emission is also a direct indicator of
AGN activity. Since the \Bootes\ field has multiwavelength coverage
that permits using all these, we investigate all of them in this
section.

\begin{figure*}
\epsscale{0.75}
\figurenum{3}
\plotone{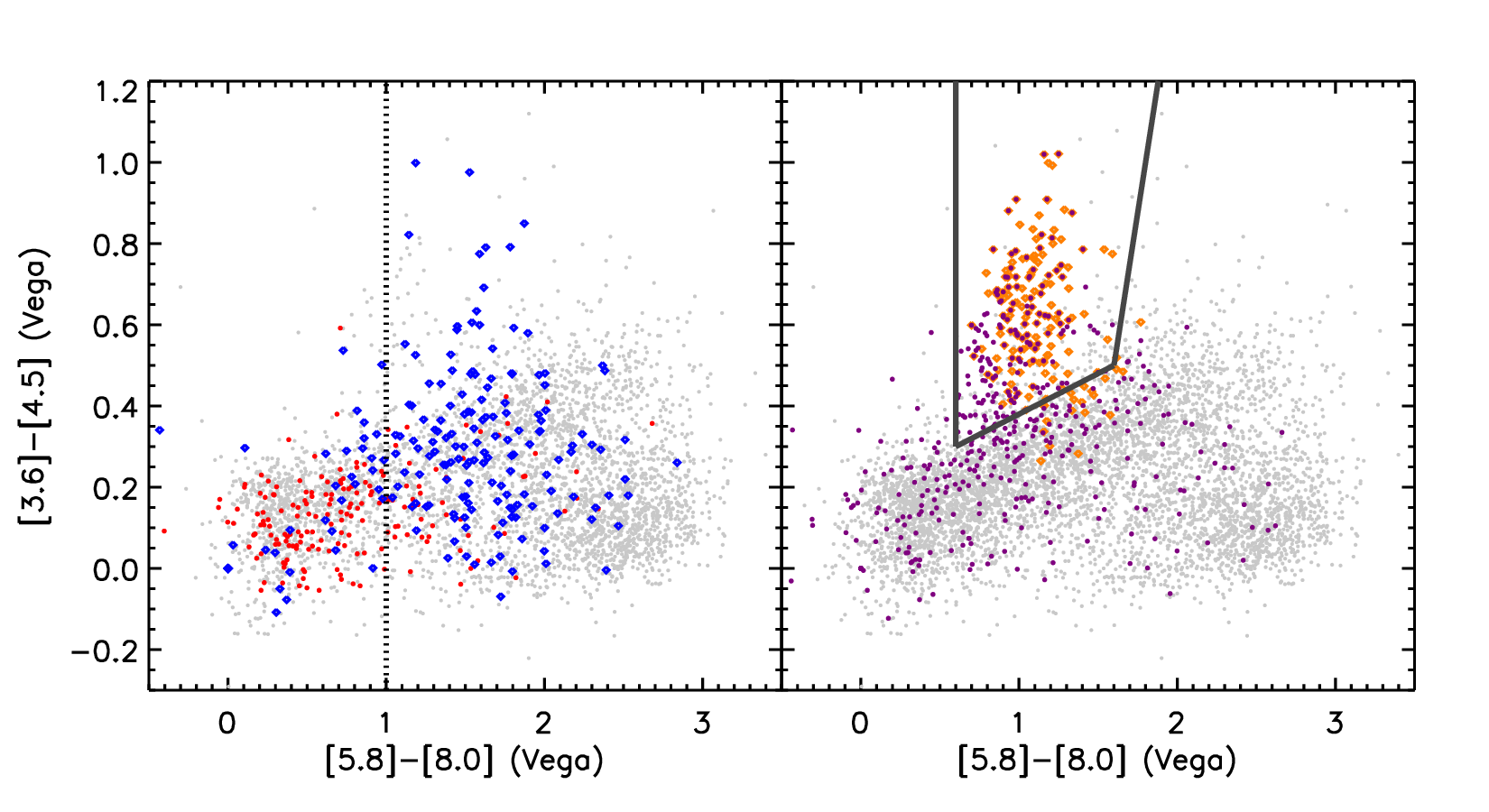}
\caption{Mid-IR properties of AGNs in our sample. Left: the
  [3.6]$-$[4.5] vs. [5.8]$-$[8.0] color-color diagram separates objects
  with and without aromatic emissions at an empirical boundary of
  [5.8]$-$[8.0] $\sim 1$ \citep[vertical line,][]{Brand09}, with
  aromatic-emitting objects on the redder [5.8]$-$[8.0] side. Blue and
  red dots represent optically selected AGN with and without 24
  $\micron$ detection, respectively. Right: the same diagram showing
  the \citet{Stern05} AGN selection criteria (trapezoidal box),
  power-law selected AGN (orange diamonds), and X-ray selected
  AGN (magenta dots). A majority of optically selected AGN that
  emit strongly at 24 $\micron$ are aromatic-emitting objects and
  their MIR emission is likely dominated by star-forming activity. On
  the other hand, a majority of X-ray selected AGN share the same
  locus as power-law selected AGN, indicating that their AGN component
  could contribute significantly at 24 $\micron$.}
\label{AGN_diagplot_IRAC}
\end{figure*}

\subsubsection{Optical Emission Line AGN Diagnostics}\label{sec:IR_LF_AGN_BPT}
In addition to identifying the broad-line AGN (type 1), our
spectroscopic survey allows us to identify narrow-line AGN (type 2) in the 
sample using the BPT diagram. We require that all four relevant lines
(H$\alpha$, H$\beta$, [OIII], and [NII])
are detected at S/N $ >3$. Thus, the BPT method can only be used for
redshift of $z < 0.35$ because at higher redshifts the H$\alpha$
emission line shifts beyond the red limit of the spectral  
range covered by Hectospec. We identify AGN on the BPT diagram
(Figure \ref{AGN_BPT_Venn}, left) as objects with line ratios
above the theoretical limit \citep[hereafter the Kewley
  line;][]{Kewley01,Kewley07} of ratios that star-forming regions could
produce. The Kewley line was estimated by considering all possible
combinations of star-forming galaxy spectra based on the {\tt
  Starburst99} \citep{Starburst99} models. Objects above the Kewley
line will be referred to hereafter as {\it optically selected AGN}. We
identified 421 optically selected AGN at $0.0 \leq z \leq 0.35$, 46\% of
which (192/421) have $F_\nu(24~\micron) > 0.27$ mJy. 

However, we need to determine whether the 24 $\micron$ emission from
optically selected AGN is actually dominated by the AGN. Our
$1\farcs5$ diameter spectroscopic fiber would cover a distance of 6
kpc at the median redshift of the AGES sample ($z = 0.26$) and the
fiber would be placed at the center of the target galaxy. Thus our BPT
classification would be sensitive to the nuclear flux from the AGN but
could miss the emission lines contributions from star-forming regions
outside the fiber. In the case that star formation 
dominates the global optical emission but an AGN dominates the emission
within the fiber, the global 24 $\micron$ flux could in fact be
dominated by the emission from star-forming activity and excluding such
objects would lead to an underestimation of the star formation rate. 

We therefore investigate the origin of mid-IR emissions from the
optically selected AGN by studying the IRAC colors of these AGN. Using
IRAC photometry from the IRAC Shallow Survey (6$^{\prime\prime}$
aperture, Section \ref{sec:data}), we study the [3.6 $\micron$] $-$ [4.5
  $\micron$] versus [5.8 $\micron$] $-$ [8.0 $\micron$] color-color
diagram (hereafter, [3.6]$-$[4.5] and [5.8]$-$[8.0], respectively) for
optically selected AGN (Figure \ref{AGN_diagplot_IRAC}, left). The
color-color diagram can be used to identify MIR aromatic emission
characteristic of star formation activity \citep{JDSmith07, Shi07},
because the strongest aromatic features at 6.2, 7.7, and 8.6 $\micron$
fall into the 8.0 $\micron$ band (channel 4; $\simeq$ 6.2 $-$ 10
$\micron$) at redshifts $0.0 \leq z \leq 0.5$. This behavior results
in a redder [5.8]$-$[8.0] color for objects with MIR aromatic emission,
which also implies that the MIR emission of these objects is 
dominated by star formation. The empirical color boundary that
separates the aromatic and non-aromatic region is [5.8]$-$[8.0] $\simeq 1$
\citep{Brand09}. We found optically selected AGN distributed across
the full range of [5.8]$-$[8.0] color. However, if we further separate
optically selected AGN into those detected at 24 $\micron$ and those
that are not, they split into two locales. Optically selected 
AGN without 24 $\micron$ detection are relatively blue in [5.8]$-$[8.0]
color while more than 70\% of those with 24 $\micron$ detection
congregate in the aromatic locus. The bluer MIR locus of optically
selected AGN without 24 $\micron$ detection is consistent with the
Rayleigh-Jeans tail of an old stellar population's photospheric
emission that peaks at 1.6 $\micron$. This suggests the source of 24
$\micron$ emission in an optically selected AGN is likely to be
star-forming activity, not the AGN.

\subsubsection{X-ray Detections}\label{sec:IR_LF_AGN_Xray}
We have also searched for AGN detected in the 5 ks X-ray survey of the
\Bootes\ field \citep[XBo\"{o}tes;][]{Murray05}. We define an {\it
  X-ray selected AGN} as any source with two or more X-ray counts and
a $> 25$\% probability based on the Bayesian matching method described
by \citet{Brand06}. Since XBo\"{o}tes is designed to have a large
contiguous area and a shallow flux limit, X-ray identification based
on the survey will only identify the strongest AGN. The X-ray
luminosities of these sources are typically brighter than $10^{42}$ ${\rm
  erg}s^{-1}$ \citep{Hickox09} and hence they are unlikely to be
dominated by star-forming activity. Although it is not possible to
infer directly the AGN contribution to the 24 $\micron$ flux from the
strength of the X-ray emission, strong AGN are known to emit
significantly in the MIR \citep[e.g.,][]{Barmby06}. We found 498 X-ray
selected AGN at $0.0 \leq z \leq 0.65$ and 175 of them are at the
redshift range where we construct the LF ($0.05 < z < 0.65$), are in
the main galaxy sample, and are detected at 24 $\micron$.

\begin{figure*}
\epsscale{0.65}
\figurenum{4}
\plotone{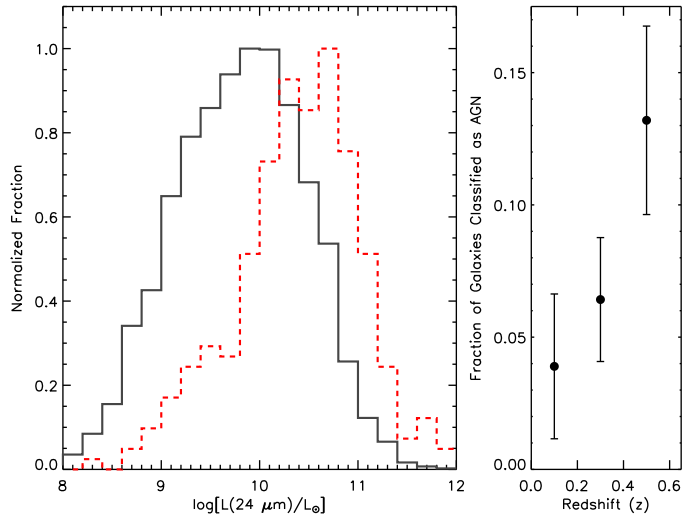}
\caption{24 $\micron$ luminosity distribution of the star-forming
  and AGN sub-samples at $0.05 \leq z \leq 0.65$. Left: the
  rest-frame \LMIPS\ distribution of 3,759 star-forming
  galaxies (solid histogram) and 288 AGNs (dashed
    histogram) identified by their characteristic power-law SED
  (based on power-law fitting and IRAC colors) and X-ray detections
  that pass all targeting and quality cuts. Right: the fraction
  of galaxies identified as AGN in our sample rises with redshift at
  $z \lesssim 0.6$. This is partly due to the fact that AGNs are
  intrinsically more luminous, as shown in the left
    panel. ULIRGs (galaxies with $L(IR) > 10^{12}$ \Lsun\ or $\LMIPS
  > 1.5 \times 10^{11}$ \Lsun) are dominated by galaxies exhibiting
  AGN activity that are excluded from our LFs of star-forming galaxies.}
\label{frac_galagn_L+z}
\end{figure*}

\subsubsection{MIR AGN Diagnostics}\label{sec:IR_LF_AGN_IRAC}
AGN with significant MIR emission can be identified most directly by
applying MIR selection methods. MIR photometry can be used to identify
luminous AGN due to their characteristic power-law SED rising from the
optical wavelengths to the IR. This power-law SED is due to the
combination of thermal emission of warm and hot dust components and
non-thermal emission of the nuclear region \citep{Neugebauer79,
  Elvis94, RiekeLebof81}. AGN can be identified by fitting a power
law, $F_\nu = \nu^{\alpha}$, to all four bands of IRAC photometry
covering 3.6 - 8 $\micron$ \citep{Eisenhardt04} with $\chi^2$
minimization. We require an object to be detected in all four bands at
above 5-$\sigma$ for power-law  fitting. {\it Power-law selected AGN}
are defined as galaxies exhibiting power-laws with $\alpha$ ranging
from $-0.5$ to $-2.0$ \citep{Ivezic02,AH06,Donley07}. The power-law
method identified 169 AGN. Among these AGN, 144 are detected at 24
$\micron$ (85\%), which is the highest fraction among the AGN
classification methods. This suggests that the 24 $\micron$ emission
of power-law selected AGN is an extension of their power-law AGN SED
to longer wavelengths rather than from the warm dust characteristic of
star-forming galaxies. Therefore the power-law selection of AGN
provides a good basis to exclude AGN from our 24 $\micron$ star-forming
galaxy sample. However, the signal-to-noise requirement on the IRAC
photometry for power-law fitting affects the completeness of power-law
AGN selection. Among these 169 AGN, 73 are at $0.05 < z < 0.65$ and pass
all targeting and quality cuts, and are detected at 24 $\micron$.

Another manifestation of power-law SED of AGN is their locus on MIR
color-color diagrams (Section \ref{sec:IR_LF_AGN_BPT}). Again, we consider
the [3.6]$-$[4.5] vs. [5.8]$-$[8.0] color-color diagram. AGN (mostly type
1) exhibit power-law SEDs and this results in a distinct locus that is
significantly redder in the [3.6]$-$[4.5] color than both early type and
star-forming galaxies \citep[see Figure 1 of][]{Stern05}. The locus of
early type galaxies, which also exhibit power-law SEDs but 
with $\alpha > -0.5$, would lie blueward in both color indices
(lower-left locus on the diagram) because, as mentioned earlier, their
SEDs in the NIR are dominated by the Rayleigh-Jeans tail of old stellar
populations. \citet{Stern05} empirically define the AGN locus on the
color-color diagram (the ``Stern wedge'', Figure
\ref{AGN_diagplot_IRAC}, right). AGN selected based on this
color-color criteria 
(hereafter {\it IRAC color selected AGN}) are defined by galaxies with
$([5.8] - [8.0]) > 0.6$ and $([3.6] - [4.5]) > 0.2([5.8] - [8.0]) +
0.18$ and $([3.6] - [4.5]) > 2.5([5.8] - [8.0]) - 3.5$. We found 354
IRAC color selected AGN in the redshift range $0.0 \leq z \leq 0.65$, 80\%
(282/354) of which have 24 $\micron$ detections. A majority (86\%,
146/169) of the power-law selected AGN are also selected by the
\citet{Stern05} criteria. The reverse is not the case: only 41\%
(146/354) of IRAC color selected AGN exhibit identifiable power-law
characteristics. However, Figure \ref{AGN_diagplot_IRAC} (right)
illustrates that there are a few galaxies in the power-law selected
AGN locus but not identified as AGN by the MIR power-law
fitting. These galaxies satisfy the \citet{Stern05} criteria, which
suggests that adopting the \citet{Stern05} AGN selection criteria
helps augment the completeness of power-law AGN selection in our
case. Among these 354 IRAC color selected AGN, 159 are at $0.05 < z <
0.65$ and pass all targeting and quality cuts, and are detected at 24
$\micron$.

As another test for MIR-emitting AGN, we used the behavior of the IRAC
[5.8] $-$ [3.6] color. For $z \ge 0.2$, both bands should be on the
Rayleigh-Jeans tail of the photospheric SED without substantial
contamination by infrared excess emission from normal star formation
(see, e.g., the \citet{Rieke09} SED templates). However, the warm
dust around an AGN should appear as an anomalously red color. Many of
the objects identified in this way are still dominated by aromatic
emission, as judged by their full set of IRAC colors. However, a small
fraction (21 objects at $0.05 < z < 0.65$) may have IR excesses
associated with AGN that are not identified by other methods. These 21
objects amount to only 7\% of our AGN sub-sample. Without other
independent AGN signatures we cannot determine whether the 24 micron
emission of these galaxies is dominated by AGN. Our results are
independent, within the errors, of whether we include or exclude these
sources from our star-forming galaxy sample.

\subsubsection{Summary of Multiwavelength AGN
  Selection}\label{sec:IR_LF_AGN_conclusion} We found that optical
line diagnostics (the BPT diagram) select AGN 
that frequently have 24 $\micron$ fluxes dominated by emission from
star formation (Section \ref{sec:IR_LF_AGN_BPT}) and therefore that these
diagnostics should not be used as a basis to exclude AGN from the MIR
star-forming galaxy sample. Methods based on power-law characteristic
of AGN (power-law SED selection \citep{Donley08} and color criteria, e.g. 
\citet{Stern05}) yield AGN samples more relevant to our 24
$\micron$ study (Section \ref{sec:IR_LF_AGN_IRAC}). On the [3.6]$-$[4.5]
vs. [5.8]$-$[8.0] color-color diagram, X-ray selected AGN occupy regions
encompassing the locus of power-law selected AGN and extend toward the
bluer [5.8]$-$[8.0] and [3.6]$-$[4.5] color regions
\citep{Cardamone08,Gorjian08}, indicative of older stellar
populations, suggesting that X-ray selection helps augment the
completeness of AGN selection by the MIR selection methods. Therefore
we combine MIR selection of AGN with X-ray selection to create a list
of AGN to be excluded from our 24 $\micron$ LF study.

The Venn diagram (Figure \ref{AGN_BPT_Venn}, right) illustrate the
significant overlapping of AGN selected by X-ray, Power-law fitting,
and the \citet{Stern05} criteria. At $0.0 \leq z \leq 0.65$, we find a total
of 712 AGNs, 288 of which are detected at 24 $\micron$, pass all
targeting and quality cuts, and are in the redshift range $0.05 \leq z
\leq 0.65$ where we use to construct LFs. Figure \ref{frac_galagn_L+z}
illustrates the fraction of AGN as a function of 24 $\micron$
luminosity and redshift. Note that objects with $\LMIPS > 10^{11}$
\Lsun\ are mostly AGN. Most of the trend in the AGN fraction with
redshift seen in Figure \ref{frac_galagn_L+z}) are a consequence of the
increasing fraction of AGN at higher \LMIPS. Higher redshift bins are
dominated by higher luminosity galaxies because of the survey flux
limit and hence will show higher AGN fractions.  

Excluding these 288 AGNs from the 4,047 galaxies at $0.05 \leq z \leq
0.65$ that are detected at 24 $\micron$, we obtain a sub-sample of
3,759 star-forming galaxies to construct the LFs.

\subsection{Methodology}\label{sec:IR_LF_Method}
We estimate our LF at redshift $0.05 \leq z \leq 0.65$ using the $1/V_{max}$
formalism \citep{Schmidt68,Huchra73}, and the local 24 $\micron$ LF
(hereafter {\it local LF} or {\it LLF}) using the parametric maximum
likelihood estimator \citep[MLE; e.g.,][]{STY79, Yahil91,
  Marshall83}. The local LF is then evolved to fit the higher redshift
1/$V_{\rm max}$ LFs using a $\chi^2$ minimization to determine the
evolution of the LF as a function of redshift
(Section \ref{sec:IR_LF_evolparam}).

The $1/V_{max}$ estimator is non-parametric in the sense that it does
not assume a functional form for the LF. However, it is sensitive to
density variations due to large scale structures. The large areal
coverage of the NDWFS helps to mitigate this problem. But at the
lowest redshifts that we observed, even a 9 sq. deg field covers a
sufficiently small volume to show significant cosmic variance. The
most prominent example is a $\sim$50\% overdensity near $z \simeq
0.15$ (see Figure \ref{Fig_Lz}).

We construct the $1/V_{max}$ LF by first computing a spectroscopic
selection function that takes into account the parent catalog's
incompleteness due to imaging and targeting issues. These issues
include sparse sampling due to priority conflicts in Hectospec fiber
assignment, incompleteness due to fiber allocation limits (such as the
proximity limit), and redshift extraction failure
(Section \ref{sec:data_AGES}). The selection function is then applied to
each galaxy as a weight, $w_i$, that corrects for incompleteness. The
$V_{max}$ in our analysis is taken to be the volume enclosed by the
maximum redshift that allows an object to be included in both our
I-band (spectroscopic targeting) and 24 $\micron$ samples. That is, we
take the smaller of the redshift at which $F_\nu(24~\micron)$ would
drop below 0.27 mJy or the redshift at which the object will have
$I_{C} > 20$ mag. We estimate the $1/V_{max}$ LF in four independent
redshift bins ranging from redshift $z = 0.05$ to $z = 0.65$ with a
bin width of 0.15. The lowest redshift region (e.g. $z < 0.05$) was
excluded to avoid potentially unrealistic number densities due to the
very small volume and number of galaxies and also to minimize the
effect of photometric bias due to nearby, large galaxies that extend
beyond our 24 $\micron$ PSF. This redshift cut excludes only 82
objects (2\% of the sample). In each bin, we take $V_{min}$ to 
be the volume enclosed by the $z_{max}$ of the adjacent lower redshift
bin.

The MLE method was used to determine the shape of the local 24
$\micron$ luminosity function (LLF). MLE requires no binning of data
and is insensitive to density variations, which is ideal for the
lowest redshift regime where large scale structures affect the
$1/V_{max}$ LF the most. We parameterize the LF with the double-power
law proposed by \citet{Yahil91}. 
\begin{equation}\Psi(>L) = C\left(\frac{L}{L_*}
\right)^{-\alpha}\left(1 + \frac{L}{L_*}\right)^{-\beta},
\end{equation}
which corresponds to the differential luminosity function
\begin{equation}
\Phi(L) = \left(\frac{\alpha}{L}+\frac{\beta}{L+L_*}\right)\Psi(>L).
\end{equation}
\noindent This functional form is commonly used to describe the MIR
luminosity function \citep{Shupe98, Xu98, Xu2000, Sanders03,
  Magnelli09}. The \citet{Xu2000} template was also used by
\citet{LeFloch05} to describe MIR LF evolution to $z \sim 1$. It
should be noted that the turnover of the Schechter function
\citep{Schechter76} frequently used for optical LFs is too steep to
fit the IR LF \citep{RiekeLebof86, Yahil91, LeFloch05}.

We determine the parameters by maximizing the likelihood
function over the parameter space of the bright-end slope, $\alpha$,
the faint-end slope, $\beta$, and the characteristic luminosity,
L$_*$, using the sample members in the redshift range of $0.05 \leq z \leq
0.25$. The normalization factor, $C$, has to be determined separately
(since the factor cancels and does not appear in the likelihood
function) by matching the integrated number density expected from MLE
to that observed in the 1$/V_{max}$ LF of the same sample. We
constructed the LLF for both star-forming and all-galaxy samples.

For the purpose of constructing a local luminosity function, the
redshift range $0.05 \leq z \leq 0.25$ cannot be considered purely
local. Especially with a population with the strong luminosity
evolution of IR galaxies, the L$_*$ estimate from such a wide redshift
range would inherently be brighter than the true local L$_*$. To
illustrate this, if we assume that the luminosity of IR galaxies
evolves as $(1+z)^{3.3}$ and the median redshift of the sample used
for the LLF is $z = 0.15$, we would expect a 60\% luminosity increase
from the actual L$_*(z = 0)$. There are a number of possible
approaches to solve this problem. The first is to fit the luminosity
evolution as a parameter in the MLE simultaneously while constraining
the shape parameters. In practice such a fit would be difficult due to
the additional free parameters  and the degeneracy between luminosity
and density evolution. One might need to fix some shape parameters
(such as the faint-end slope) to achieve a plausible fit. The second
approach would be to limit the redshift range of the sample to
estimate the LLF. However, reducing the redshift bin size would
compromise the constraints on the bright-end slope; we need to
estimate the LLF using a sample selected to $z \leq 0.25$ to encompass
enough volume to find rare, bright galaxies. Others have faced the same
problem. The \citet{Shupe98} and \citet{Xu2000} LLFs contain few
bright galaxies as far as $z \sim 0.2$. Although their median redshift
is much closer to local, the luminosity evolution could result in a
more luminous L$_*$ than the true local value. Another way to limit
the redshift range to the local volume would be to rely on an MIR
sample from all-sky IR surveys with shallow limiting fluxes such that
the sample is limited to very nearby galaxies. But since all-sky
surveys lack complete spectroscopic redshift follow-up and usually
have to compile redshifts from several catalogs, the spectroscopic
selection function of the combined sample may not be well
understood. Robustly constraining a true local L$_*$ is important
because the local L$_*$ is often used to constrain the evolution of
the LF at higher redshifts, such as in \citet{LeFloch05}, who used the
\citet{Sanders03} local L$_*$ derived from the IRAS 60
$\micron$-selected sample. 

Our solution to this problem is to take advantage of the evolution
information in our full sample, an option not available to studies
limited to $z \lesssim 0.2$. The evolution constraint from the higher
redshifts can be used to constrain the evolution and extrapolate L$_*$
back to the true value at $z = 0.0$ (Section \ref{sec:IR_LF_evolparam}). 

\begin{figure}
\epsscale{1.20}
\figurenum{5}
\plotone{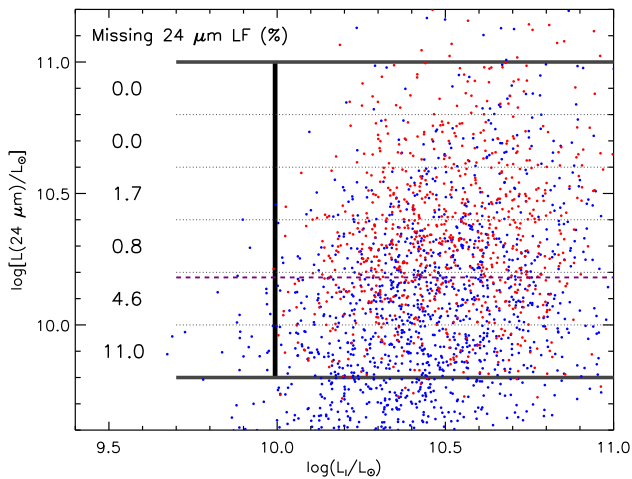}
\caption{24 $\micron$ and I-band bivariate luminosity functions
  (BLF) were constructed to quantify the effects of incompleteness at
  24 $\micron$ of MIR-bright, optically-faint galaxies (the {\it bivariate
  selection bias}). The BLF construction to quantify the bivariate
  selection bias for the sub-sample at redshift $0.35 \leq z \leq 0.5$ is
  illustrated in this figure. The blue dots illustrate the sample
  observed at $0.20 \leq z \leq 0.35$ evolved in both I-band and 24 $\micron$
  luminosities to match the distribution at $0.35 \leq z \leq 0.5$ (shown in
  red dots as a comparison). The solid vertical line is the
  luminosity, corresponding to the I-band limiting
  magnitude ($I=20$) at $z = 0.35$, below which a galaxy will be
  excluded from our $0.35 \leq z \leq 0.5$ sub-sample. The values indicated
  in the missing fraction column are the percent of the 24 $\micron$ LF
  potentially excluded due to the bivariate selection bias (fraction
  of 24 $\micron$ LF contributed by objects represented by blue dots
  on the left of the thick vertical line) in each \LMIPS\ bin
  (separated by the horizontal dotted lines). The 
  horizontal dashed line indicates the luminosity corresponding
  to the 24 $\micron$ completeness limit (0.27 mJy) at $z = 0.50$,
  above which 24 $\micron$ sub-sample at $0.35 \leq z \leq 0.5$ is
  complete. We quantified the missing LF due to this effect to be less
  than 2\% at \LMIPS\ above the completeness limit,
  indicating that our spectroscopic targeting method did not exclude
  significant numbers of MIR-bright, optically-faint galaxies.}
\label{LF_bivar_LF}
\end{figure}

\begin{figure*}
\epsscale{0.675}
\figurenum{6}
\plotone{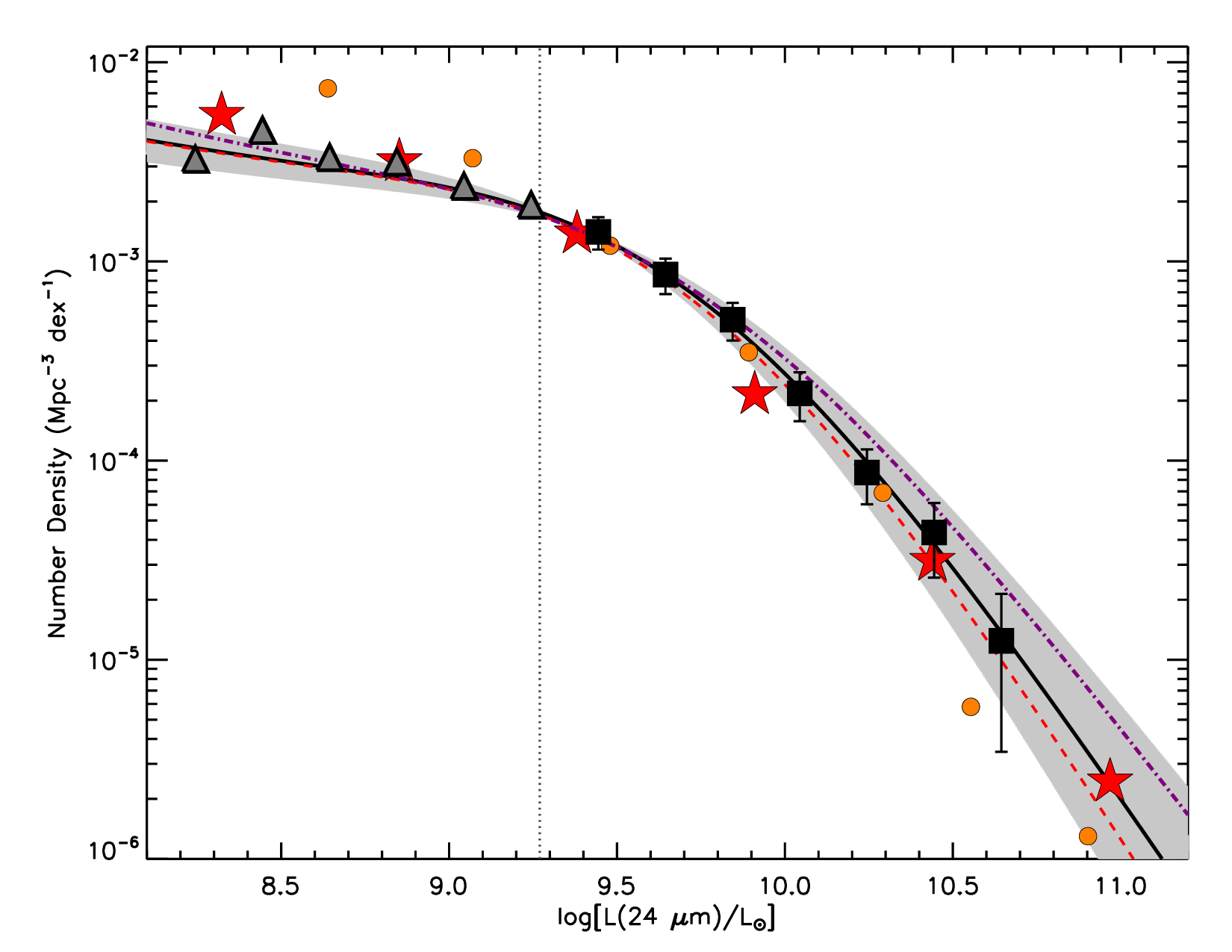}
\caption{Mid-IR local luminosity function (LLF). We construct the
  LLF for the all-galaxy (solid black line) and star-forming
  (dashed red line) samples. The $1/V_{max}$ LLF points are
  shown in black squares above the luminosity corresponding to the 24 $\micron$
  completeness flux limit at the upper limit of redshift range where
  we select the sub-sample to construct the LLF (0.27 mJy at $z =
  0.25$, vertical line) and lower-limits (upward triangles) below
this completeness threshold. Note that this 
  completeness limit does not affect the maximum likelihood estimator
  (MLE) LLF. Uncertainties in the bright and faint-end slopes of the
  MLE LLF derived from the jackknife analysis are illustrated by the
  shaded area. The mid-IR LLFs from \citet{Rush93,Shupe98} and
  \citet{Sanders03} are corrected for bandpass differences and shown
  in circles, dot-dashed line, and stars, respectively. The
  consistency of our LLF shape with previous work shows that it is a
  reliable local anchor point to study the evolution of the LF.}
\label{LF_LLF}

\epsscale{0.7}
\figurenum{7}
\plotone{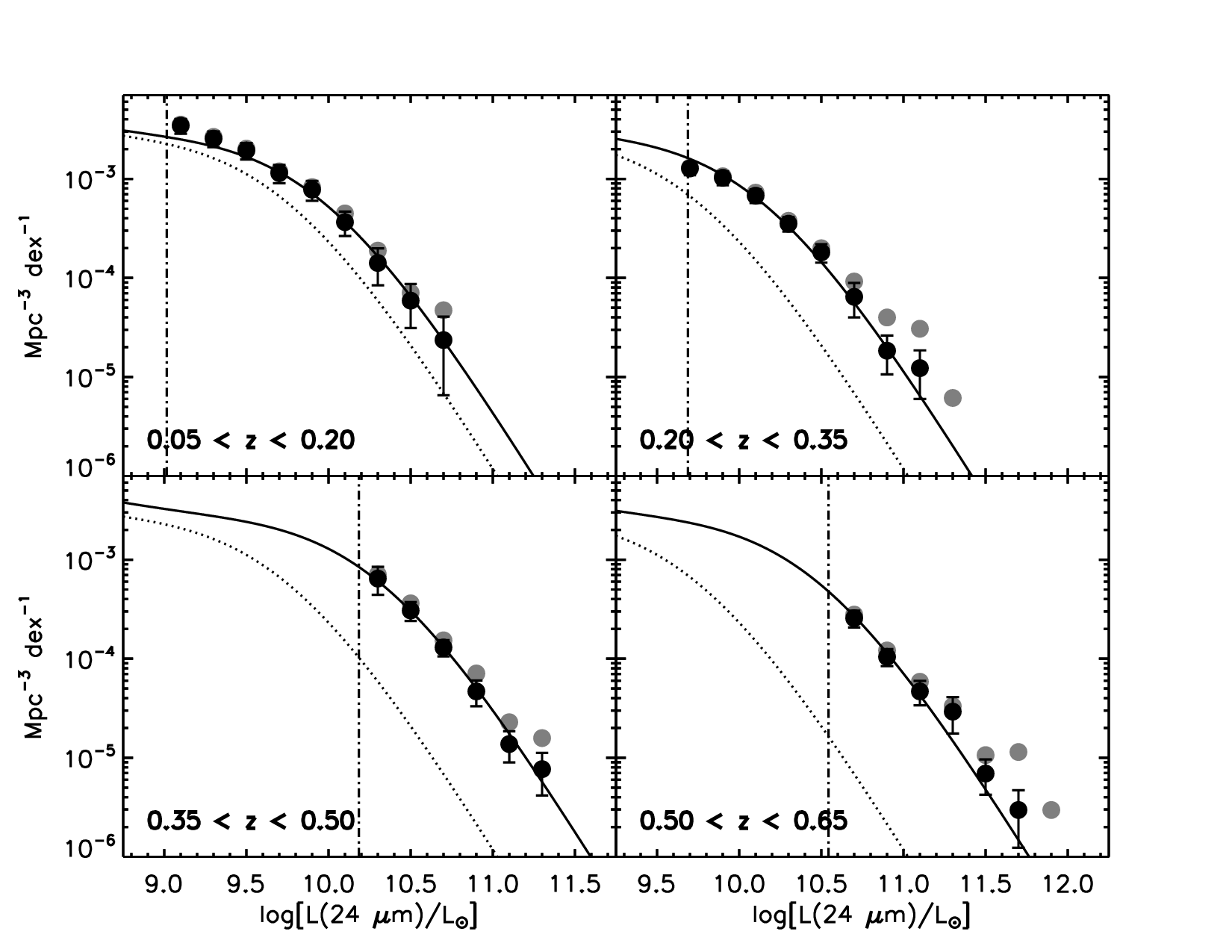}
\caption{Evolution of the 24 $\micron$ luminosity function (LF) of
  star-forming galaxies (black circles) and all galaxies (gray circles,
  error bars omitted for clarity) for four redshift bins between $0.05
  \leq z \leq 0.65$. The dotted LF is the local 24 $\micron$ LF of
  star-forming galaxies determined by the maximum likelihood estimator
  and the solid line in each panel is the local LF evolve to fit the
  star-forming galaxy $1/V_{max}$ LF of the corresponding redshift bin
  (see Sections \ref{sec:IR_LF_evolparam} and
  \ref{sec:results_LF_evol}). The vertical dot-dashed lines 
  indicate the luminosity corresponding to the 24 $\micron$
  completeness flux at the upper limit of the redshift bin,
  above which our sample is complete. We note that the 1/$V_{\rm max}$ LF
  in the lowest redshift bin could be affected by the overdense region
  near $z \simeq 0.15$ (see Figure \ref{Fig_Lz}). The strong evolution
  of the LF can be described by pure-luminosity evolution.}
\label{LF_allbins}
\end{figure*}

\begin{deluxetable*}{ccccc}
\tablecaption{Parameters of the Mid-IR Local Luminosity Function in
  Comparison with Other Works} 
\tablehead{Reference\tablenotemark{a} & $\alpha$ & $\beta$ & $L_{*}(24~\micron)$ & C\\
 &  &  & (10$^9 $L$_{\odot})$ & (10$^{-3} {\rm Mpc}^{-3}$)\\}
\startdata
All Galaxies & 0.36 $\pm$ 0.04 & 2.17 $\pm$ 0.35 & 4.07 $\pm$ 0.67 & 1.2 $\pm$ 0.8\\
Star-Forming  & 0.37 $\pm$ 0.04 & 2.36 $\pm$ 0.41 & 4.27 $\pm$ 0.71 & 1.2 $\pm$ 0.8\\
\citet{Shupe98} & 0.437 $\pm$ 0.032 & 1.749 $\pm$ 0.067 & 4.47 $\pm$ 0.40 & 1.0 $\pm$ 0.6\\
\citet{Xu98} & 0.47 $\pm$ 0.13 & 2.20 $\pm$ 0.13 & 6.40 $\pm$ 2.25 & 0.9 $\pm$ 0.5\\
\citet{Sanders03} & 0.6 $\pm$ 0.1 & 2.2 $\pm$ 0.1 & 4.4 & 0.8\tablenotemark{b}
\enddata
\tablenotetext{a}{Results from other studies are corrected for their
  bandpass differences to reflect estimates at 24 $\micron$ using the
  \citet{Rieke09} SED library. Redshift
  ranges for each study are the following. This study: $0.05 \leq z
  \leq 0.25$, median redshift of 0.17 in both samples, with L$^{*}$
  evolve back to $z = 0.0$ (for both all-galaxy and star-forming
  sample) simultaneously with the evolution fit (see
  Section \ref{sec:IR_LF_evolparam}); \citet{Shupe98}: $0.001 \leq z \leq
  0.07$; \citet{Xu98}: $0.02 \leq z \leq 0.22$; \citet{Sanders03}:
  median and mean redshifts of 0.0082 and 0.0126, respectively.}
\tablenotetext{b}{\citet{Sanders03} do not report their normalization
  explicitly but a power-law break fit to their 1$/V_{max}$ LLF (see
  their Table 6) suggests a value of $8 \times 10^{-4}$ ${\rm
    Mpc}^{-3}$.}
\label{table_comparellf}
\end{deluxetable*}

\subsection{Bivariate Selection Bias}\label{sec:IR_LF_Bivar}
Since AGES spectroscopy was targeted on a flux-limited I-band parent
catalog, our sample could be affected by biases created by excluding
24 $\micron$-bright, optically faint galaxies. This {\it bivariate
  selection bias} may result in a sample that systematically excludes
galaxies with optically obscured star formation. 

We quantify the possible bias by estimating the fraction of the 24
$\micron$ number densities excluded because their contributing sources
fall below the I-band flux limit. This can be done by constructing a
24 $\micron$ LF as a function of I-band luminosity ({\it bivariate
  luminosity function, or BLF}) of a model sample constructed by
evolving a sample at lower redshift, where it is complete to a fainter
I-band luminosity, to a higher redshift bin of interest. To illustrate this
experiment, consider the BLF in Figure \ref{LF_bivar_LF}. In this case,
we are investigating the bivariate selection effect for $0.35 \leq z \leq
0.50$. We construct a model sample by evolving a sample from $0.20 \leq z
\leq 0.35$ (which is complete to deeper luminosity) in both I-band and 24
$\micron$ to construct a BLF at $0.35 \leq z \leq 0.50$. As illustrated in
Figure \ref{LF_bivar_LF}, some objects would fall below the luminosity
corresponding to the I-band flux limit ($I=20.0$) at the low-redshift
edge of the bin ($z = 0.35$) and would therefore be excluded from our
sample, resulting in a fraction of the missing 24 $\micron$ LF. We found,
however, that this fraction is small ($< 2$\%) in this redshift bin, for 
24 $\micron$ luminosities above our 24 $\micron$ completeness flux.

To model the evolution of $L(I)$, we refer to \citet{Willmer06} and
\citet{Brown07} who found that the optical characteristic magnitudes
brighten by $\sim 1.3$ mag per unit redshift, which would correspond to
a luminosity evolution law $L \propto (1+z)^{1.6}$ at $0 \leq z \leq
1$. For the 24 $\micron$ luminosity evolution model, we adopted $L
\propto (1+z)^{3.3}$ based on \citet{Pozzi04,LeFloch05} and
\citet{PPG05}.

We repeated this test for the other two higher redshift bins and found
that the bivariate selection bias is negligible in the $0.20 \leq z \leq
0.35$ bin; none of galaxies is excluded from the sample in this
redshift bin due to bivariate selection bias. Only in the highest
redshift bin, $0.50 \leq z \leq 0.65$, does the missing fraction
become more significant. In this bin we found the missing fraction in
the \LMIPS\ bins of $10^{10.7}$, $10^{10.9}$, and $10^{11.1}$
\Lsun\ to be 8.7\%, 4.4\%, and 0.0\% respectively.

The bivariate selection biases are under 5\% for all but the lowest
luminosity bin of our highest redshift sub-sample. We correct for it
by by applying a correction factor based on the missing fraction
estimated for each bin of the 1$/V_{max}$ LF. This correction is
included in the LF shown in Figure \ref{LF_allbins}. In a broader
picture, this test indicates that our spectroscopic targeting method
does not systematically exclude MIR-bright, optically-faint galaxies
from the final sample.

\subsection{Evolution of the Luminosity
  Function}\label{sec:IR_LF_evolparam}  
The evolution of the 24 $\micron$ LF is quantified using two methods. The
first is to evolve the local LF template (Section \ref{sec:IR_LF_Method})
to fit the 1$/V_{max}$ LF at higher redshifts so we can estimate the
characteristic luminosity, L$_*$, and any evolution in comoving number
density as a function of redshift. The second is to determine the
luminosity in each redshift bin at which the cumulative number density
reaches a common threshold.

In the first method, we parameterize the LF evolution assuming a
functional form of
\begin{equation}\psi(L,z)=g(z)\phi\left[L/f(z)\right]
\end{equation}
\noindent where $\phi(x)$ is given by Equation (1). The terms
$f(z)=(1+z)^{\alpha_L}$, and $g(z)=(1+z)^{\alpha_D}$ describe the
evolution of the luminosity and number density with redshift,
respectively. We then take our LLF determined by the MLE method
(Section \ref{sec:IR_LF_Method}) as a local template and fit it to the
1$/V_{max}$ LF in each higher redshift bin using $\chi^2$ minimization
and the median redshift of the bin. By using the LLF as a template,
we assume a fixed LF shape across the redshift range, and focus on the
change in $L_*$ and $C$ as a function of redshift. To fit the LF
template to the 1$/V_{max}$ LF, we only consider $1/V_{max}$ data
points above the luminosity corresponding to the 80\% completeness
flux threshold (0.27 mJy) at the upper redshift limit of the
bin. Below this luminosity our sample in the corresponding redshift
bin is no longer complete. The fit is done in two ways: fitting all
redshift bins simultaneously to estimate the global evolution of the
LF at $0.0 \leq z \leq 0.6$, and fitting each bin individually to explore
any bin-specific deviations from the global law. We also integrate the
individually fitted LFs over luminosity to determine the luminosity
density, which will be used to estimate the star formation rate
density (\SFRD\, Section \ref{sec:IR_LF_SFR}). 

As mentioned in Section \ref{sec:IR_LF_Method}, the LLF is affected
by the evolution of IR galaxies within the redshift bin where the LLF
was constructed. With a known median
redshift of the LLF sample ($z = 0.17$ for both star-forming and
all-galaxy samples), we can use $\chi^2$ minimization to give the
best fit evolution law that would evolve the L$_*$ of the LLF from
$L_*(z_{median})$ back to $L_*(z = 0)$ and simultaneously evolve the
LLF with $L_*(z = 0)$ to fit the 1$/V_{max}$ LF at all higher redshift
bins. After using the global luminosity evolution constraint to evolve
L$_*$ back to $z = 0$, we can then use $L_*(z = 0)$ as a local anchor
point to quantify the luminosity evolution in each bin individually.

In the second method to quantify luminosity evolution, we determined
the luminosity at which the corresponding cumulative number density
reaches a common arbitrary threshold in each redshift bin. The {\it
  individual} fit of the LF is integrated downward in luminosity from
the bright-end until the integrated number density reaches $10^{-4}
{\rm Mpc}^{-3}$. This method robustly quantifies the luminosity 
evolution because the bright-end of our LF is very well constrained
and little affected by incompleteness. 

\subsection{Estimation of SFR}\label{sec:IR_LF_SFR}
We adopt the SFR calibration of \citet{Rieke09} to estimate the SFR
from the observed $F_\nu(24~\micron)$. Every SFR indicator using
infrared luminosity is based on the assumption that UV 
radiation from young stars is efficiently reprocessed into IR
radiation, and that the fraction of UV radiation escaping the
star-forming regions without being reprocessed is small. This escape
fraction depends strongly on the total-IR luminosity since more
luminous IR emission implies  
greater UV and optical extinction. This relationship has been studied
by comparing the $L(UV)$ and \LTIR\ of galaxies \citep{Schmitt06,
  Buat07}, who found that the UV contribution to the total luminosity
(the UV escape fraction) is about 20\% for galaxies with
\LTIR\ equivalent to $\LMIPS \simeq 10^{8.65}$
\Lsun and decreases rapidly to 8\% and 2.5\% at $\LMIPS \simeq
10^{9.6}$ \Lsun\ and $\LMIPS \simeq 10^{10.1}$ \Lsun,
respectively. Therefore the calorimetric condition can be assumed for
our normal star-forming galaxies, whose \LMIPS\ are in the range of
$10^{8} - 10^{10}$ \Lsun, although the error will increase between
$10^8$ and $10^{8.65}$ \Lsun. This is relevant only at the lowest
redshifts. We note that the \citet{Rieke09} 24 $\micron$ SFR
calibration has a scatter of $< 0.2$ dex, but a relative trend such as
the \SFRD\ evolution law will not be affected by this scatter so long
as the sample is large. The luminosity density that we use to estimate
the star formation rate density (\SFRD) is calculated by integrating
the fitted LFs (Section \ref{sec:IR_LF_evolparam}) in each redshift bin
over luminosity from $10^6$ to $10^{15}$ \Lsun.

The IR SFR indicators are calibrated with local SEDs. Employing these
indicators at higher redshifts carries an assumption of no or little
SED evolution over the relevant range of z (to 1.2). This assumption
is supported by \citet{Marcillac06}, \citet{Bavouzet08} and
\citet{Magnelli09}. There are indications that, at redshifts of $z
\sim 2$, the calibration of the observed 24 $\micron$ flux density and
SFR may shift from values deduced from local SED templates
\citep{Papovich07, Rigby08, Farrah08}. The primary cause of these
changes appears to be that the aromatic bands tend to be stronger for
a given luminosity at high redshift, so the SEDs resemble more closely
those of local galaxies with significantly smaller total infrared
luminosity. The probable cause of this behavior is that local ULIRGs
tend to have intense star formation concentrated into very optically
thick nuclear regions, whereas galaxies of similar infrared luminosity
at high redshift probably have more widely distributed star forming
regions of lower optical depth. It is unlikely that this effect
significantly alters the calibration of SFRs in this paper. First,
the effect appears to be weaker for normal star-forming galaxies and 
galaxies of LIRG-level luminosity even at high redshift
\citep{Rigby08}; the galaxies that dominate our study are of lower
luminosity still. Second, given their relatively low redshifts,
the observed 24 $\micron$ flux densities from our galaxies are not
strongly affected by the aromatic band strengths. Third, the
evolutionary state of our galaxies should resemble local galaxies at
least as much as it does those at $z \sim 2$; in fact,
\citet{Magnelli09} show that the SEDs of infrared galaxies for $0.4 \leq
z \leq 1.2$ resemble those of local galaxies reasonably closely.

\subsection{Uncertainties Analysis for the LF and
  SFR}\label{sec:IR_LF_errors} 

We use jackknife resampling to estimate the uncertainties in our
1$/V_{max}$ LFs and the LLF parameters ($\alpha$, $\beta$, $L_*$, and
the normalization factor; see Section \ref{sec:IR_LF_Method}), as well as
star formation rate density and evolution laws derived from the LFs.

The AGES survey used 15 overlapping tiles corresponding to the
Hectospec field of view. These tiles are a natural way to build 15
jackknife samples, each dropping one of these roughly 1-degree
diameter tiles and then using the variance in parameter estimates for
these 15 sub-samples to estimate the uncertainties. This procedure will
include Poisson uncertainties and sub-degree-scale cosmic variance
effects. It will underestimate uncertainties due to larger scale
correlations between galaxies.

We can estimate the large scale effects by shifting the entire LF
estimation for each redshift bin by our estimate of the total cosmic
variance and recalculating the parameters. The cosmic variance on the
scale of our survey volume ($\sim 10^6 {\rm Mpc}^3$) can be
estimated using \citet{peebles80} and by assuming the power-law
correlation function estimated from the SDSS for late-type galaxies,
$\xi = \left(r/r_0\right)^{-\gamma}$, with $r_0 = 3.6~h^{-1}{\rm Mpc}$
and $\gamma = 1.7$ \citep{Zehavi05}. In the bins of redshift $0.05 \leq z
\leq 0.20, 0.20 \leq z \leq 0.35, 0.35 \leq z \leq 0.50$, and $0.50
\leq z \leq 0.65$, we expect the uncertainties on the LF due to cosmic
variances to be 15\%, 11\%, 10\%, and 9\%, respectively \citep{Dai09}.

\section{RESULTS}\label{sec:results}
Here we will discuss our determination of the local 24 $\micron$ LF
and compare it to other studies (Section \ref{sec:results_LLF}) to establish a
baseline for our study of LF evolution to $z \sim 0.6$
(Section \ref{sec:results_LF_evol}). Next, we discuss the evolution
constraints based on the combination of our \Bootes\ field data and
the 35 $\micron$ LF evolution from \citet{Magnelli09} who focused on
$0.4 \leq z \leq 1.3$ (Section \ref{sec:results_LF_evol_to_FIDEL}) using the
FIDEL survey. The combination of these two studies lets us determine the
\SFRD\ evolution from $0 \leq z \leq 1.2$ in a unified and accurate
way. We will then present the local star formation rate density and
the star formation rate density evolution derived from the LFs
(Section \ref{sec:results_SFRD}). 

\subsection{Local 24 $\micron$ Luminosity
  Functions}\label{sec:results_LLF} 
The local 24 $\micron$ LFs for both the star-forming and all-galaxy
populations are shown in Figure \ref{LF_LLF}. Both LLFs are constructed
from a sample at $0.05 \leq z \leq 0.25$ (median redshift $z = 0.17$) and
evolved back to $z = 0.0$ using the global evolution law estimated in
Section \ref{sec:IR_LF_evolparam}. Their shape parameters are tabulated in
Table \ref{table_comparellf} along with other estimates of the LLF
using the double-power law form of equation.

The star-forming and all-galaxy LLFs have identical faint-end slopes,
but the star-forming LLF has a slightly steeper bright-end slope and a
brighter characteristic luminosity. Both differences are within the
uncertainties of the respective parameters, but nevertheless the
steeper slope of the star-forming LLF is expected because the sample
excluded the IR-luminous AGN, as illustrated in
Figure \ref{frac_galagn_L+z}. This comparison shows that the overall
impact of AGN on the LLF is small, and that any residual unidentified
AGN should have little effect on our conclusions. In Figure \ref{LF_LLF}
we present our LLF along with the bandpass and cosmologically adjusted
1$/V_{max}$ LLF of \citet{Rush93}, \citet{Sanders03} and
\citet{Pozzi04}, and the MLE LLF of \citet{Shupe98}. 

One of the most important parameters of the LLF is the value of
L$_*(z=0)$. Our L$_*$ estimates for the star-forming and all-galaxy
populations are $(4.27 \pm 0.71) \times 10^9$ and $(4.07 \pm 0.67)
\times 10^9$ $\Lsun$, respectively. We compare our results with other
works that adopt the double power-law form of LF by correcting for
cosmological assumptions and bandpasses using the \citet{Rieke09} SED
library. The \citet{Shupe98} LLF using IRAS 25 $\micron$ measurements
reports a local L$_*$ of $(4.67 \pm 0.41) \times 10^9$ $\Lsun$,
which corresponds to $(4.47 \pm 0.40) \times 10^9$ $\Lsun$ when
corrected for the difference between the IRAS 25 $\micron$ and MIPS 24
$\micron$ bandpasses. \citet{Xu98} report an ISO 15 $\micron$ local
L$_*$ of $(4.79 \pm 1.69) \times 10^9$ $\Lsun$, corresponding to
$(6.40 \pm 2.25) \times 10^9$ $\Lsun$ at 24
$\micron$. \citet{Sanders03} report a local total-IR L$_*$ of
$10^{10.5} \Lsun$, which corresponds to a 24 $\micron$ L$_*$ of $4.4
\times 10^9$ $\Lsun$. Our estimate of $L_*$ agrees well with all these
other studies. More importantly, the 
extrapolation of the LLF $L_*$ back to $z = 0.0$ using the luminosity
evolution constraint from our full sample ensures that these values
reflect an accurate value at $z = 0$ in addition to providing the
best-match LLF template to study the evolution of the LF in our
sample.

\begin{deluxetable}{ccc}
\tablewidth{0pt}
\tablecaption{The 24 $\micron$ and Total-IR Luminosities Above Which
  the Cumulative Number Density of Star-Forming Galaxies is $10^{-4}
  {\rm Mpc}^{-3}$ for Each Redshift Bin}
\tablehead{Redshift Bin & \LMIPS\ & \LTIR\tablenotemark{a} \\
 & $10^{10} \Lsun$ & $10^{10}\Lsun$}
\startdata
0.05 $\leq z \leq$ 0.20 & 1.2 $\pm$ 0.4 & 9.2 $\pm$ 2.9\\
0.20 $\leq z \leq$ 0.35 & 1.7 $\pm$ 0.4 & 13.2 $\pm$ 3.0\\
0.35 $\leq z \leq$ 0.50 & 2.5 $\pm$ 0.5 & 18.9 $\pm$ 3.7\\
0.50 $\leq z \leq$ 0.65 & 3.9 $\pm$ 0.7 & 28.1 $\pm$ 5.3
\enddata
\tablenotetext{a}{\LTIR\ is estimated using the \citet{Rieke09}
  formalism.}
\label{table_numden}
\end{deluxetable}

\begin{figure*}
\epsscale{0.7}
\figurenum{8}
\plotone{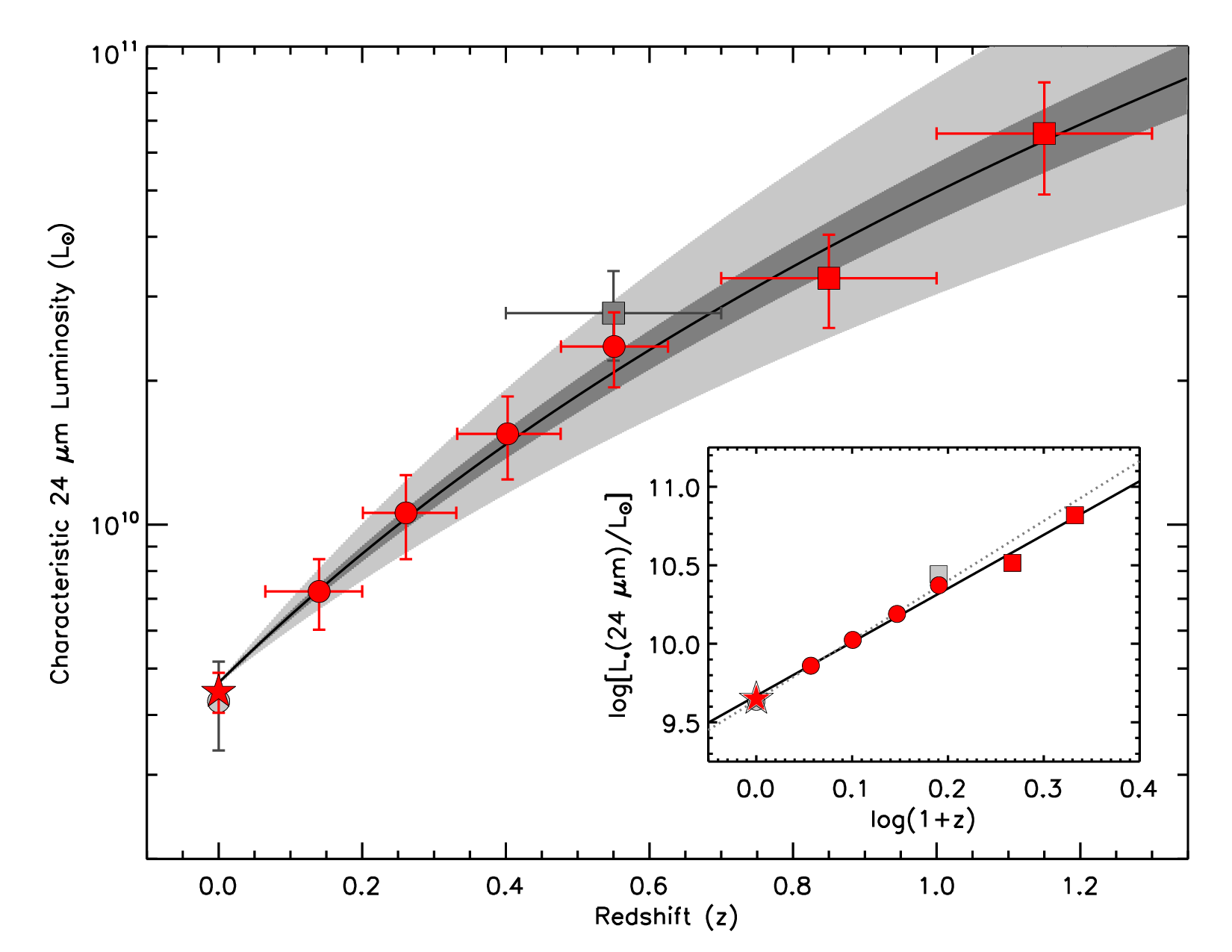}
\caption{Evolution of the characteristic 24 $\micron$ luminosity of
  galaxies, L$_*$, to $z \sim 1.2$ based on our results (red circles)
  and that of \citet{Magnelli09} using FIDEL (red squares). The light
  shaded area is the luminosity evolution law $L_* \propto
  (1+z)^{\alpha_L}$ bounded by $\alpha_L$ of 2.7 and 4.2 from the
  lowest and highest uncertainty limits of \citet{PPG05} and 
  \citet{Magnelli09}, respectively. The combined evolution law,
  $L_{*,combined} \propto (1+z)^{3.4 \pm 0.2}$, is shown as a solid
  line with the uncertainties from the combined fit shown by the dark
  shaded area. The local point for the global fit (red star) is that of
  \citet{Shupe98} converted to 24 $\micron$. Two data points were
  excluded from the fit (see details in
  Section \ref{sec:results_LF_evol_to_FIDEL}) and are shown in gray.
    Inset: the combined of luminosity evolution (solid line)
  with the same symbols as the main figure (error bars are omitted for
  clarity) and the fit (dotted line) based on the \Bootes\ data
  alone, $(1+z)^{3.8 \pm 0.3}$.}
\label{LF_Lstar_evol}

\epsscale{0.7}
\figurenum{9}
\plotone{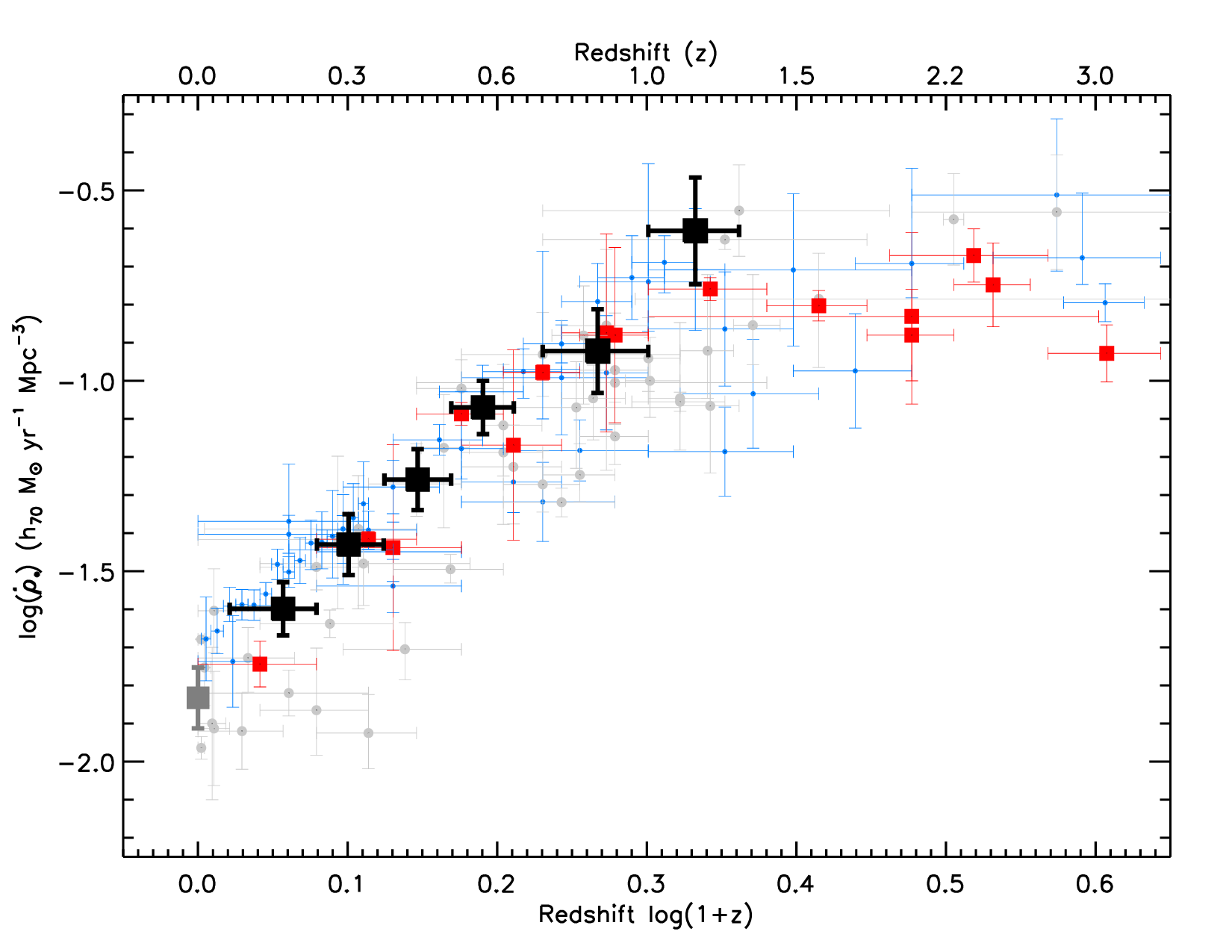}
\caption{Evolution of the star formation rate density (\SFRD) as a
  function of redshift assuming \citet{Salpeter55} IMF. Our
  \SFRD\ estimates are shown in large black squares and our
  extrapolated local point (Section \ref{sec:results_LLF}) shown in gray,
  compared to the \SFRD\ from the following. The extinction-corrected
  UV \SFRD\ from the compilation of \citet{Hopkins04} and
  \citet{HopkinsBeacom06} are shown in blue circles; IR \SFRD\ in red
  squares \citep{Hopkins04, PPG05, Reddy08}; gray dots are the radio
  (1.4 GHz), X-ray, and extinction-corrected 
  [OII], H$\alpha$, and H$\beta$ \SFRD\ \citep{Hopkins04, Seymour08,
    Smolcic09, Zhu09}. The agreement of our \SFRD\ normalization and
  those from extinction-corrected optical and UV observations confirms
  that a majority of star formation at $z < 1.2$ is obscured and can
  be directly traced by 24 $\micron$ observation.}
\label{SFR_Madauplot}
\end{figure*}

\subsection{Evolution of the 24 $\micron$ LF}\label{sec:results_LF_evol}
Strong luminosity evolution is evident in the 1$/V_{max}$ LFs over the
redshift range $0.05 \leq z \leq 0.65$ shown in Figure \ref{LF_allbins}. We
considered two evolutionary scenarios for the 24 $\micron$ LF
parameterized by Equation (3), including pure luminosity evolution and
pure density evolution. These scenarios were tested by evolving our
local 24 $\micron$ LF to fit the 1/$V_{\rm max}$ LF at higher redshift by
$\chi^2$ minimization. We cannot constrain a combined luminosity and
density evolution model because we do not sample below L$_*$ in these
higher redshifts sub-samples. The density evolution model can be
rejected on the basis of the $\chi^2$ values with high
confidence. Other studies have found that the density evolution is
small. \citet{LeFloch05} and \citet{PPG05} derived a density evolution
laws of $\phi^* \propto (1+z)^{0.7^{+0.2}_{-0.6}}$ and $\phi^* \propto
(1+z)^{1.0 \pm 0.3}$, respectively. More recently, \citet{Magnelli09}
used a stacking analysis to study the evolution of the 35 $\micron$ LF
and reported density evolution given by $\phi^* \propto (1+z)^{0.1 \pm
  0.7}$ at $z \lesssim 1.2$. This implies a density variation of about
$5\% \pm 40\%$ at $0.0 \leq z \leq 0.6$, which is comparable to the
uncertainties due to cosmic variance in the density.

Using a pure luminosity evolution model, we obtain an evolution law
given by  $L_*(24~\micron) \propto (1+z)^{3.8 \pm 0.3}$ for the
star-forming sample. This luminosity evolution is shown as a fit to the
1/$V_{\rm max}$ LF in Figure \ref{LF_allbins}. One of the evolution
descriptions of \citet{PPG05} uses pure luminosity evolution with the
\citet{Schechter76} form of LF. They report luminosity evolution of
$(1+z)^{3.6 \pm 0.3}$ for $z < 0.8$, agreeing with our result. The
evolution of the LF was also quantified by determining the luminosity
above which the cumulative number density reached $10^{-4} {\rm
  Mpc}^{-3}$; as tabulated in Table \ref{table_numden}. This
luminosity increases by more than a factor of three from a redshift of
0.05 to 0.65. We should note that while we cannot jointly constrain
luminosity and density evolution, the simple power-law model is an
adequate description of our data. Improving on it would require a
deeper sample in a comparably large area to the \Bootes\ field.

Our method to determine evolution of the LF assumes no evolution in
LF shape from the local template. The template, however, yields a
better fit at low redshift than in the higher redshift
bins. Discrepancies between the template and the data are most
visible at $0.50 \leq z \leq 0.65$ (Figure \ref{LF_allbins}). This
discrepancy could be due to an intrinsic evolution toward a shallower
bright-end slope at higher redshifts. On the other hand, it
could indicate a contribution from unidentified AGN in the higher 24
$\micron$ luminosity regime. The latter explanation is consistent
with the observation that AGN are progressively more abundant at
higher redshifts (Figure \ref{frac_galagn_L+z}) as well as from
Figure \ref{LF_allbins} itself, which shows that the bright-end slope of
the 1/$V_{\rm max}$ LF of all galaxies (including AGN) becomes shallower
to a greater extent compared to that of the star-forming galaxy LF.

\begin{deluxetable}{ccc}
\tablecaption{Evolution of the Characteristic 24 $\micron$ Luminosity,
  L$_*(24~\micron)$, and the Star Formation Rate Density, \SFRD, to $z
  \sim 1.2$}
\tablehead{Redshift & L$_*(24~\micron)$ & log(\SFRD)\tablenotemark{a} \\
 & $10^9$ \Lsun & }
\startdata
0.00 (this work)        & 4.27 $\pm$ 0.71 & $-1.83 \pm$ 0.08\\
0.00 \citep{Shupe98}    & 4.47 $\pm$ 0.40 & $-1.71 \pm$ 0.04 \\
0.05 $\leq z \leq$ 0.20 & 7.25 $\pm$ 1.12 & $-1.60 \pm$ 0.07\\
0.20 $\leq z \leq$ 0.35 & 10.6 $\pm$ 1.8  & $-1.43 \pm$ 0.08\\
0.35 $\leq z \leq$ 0.50 & 15.5 $\pm$ 2.4  & $-1.26 \pm$ 0.08\\
0.50 $\leq z \leq$ 0.65 & 23.6 $\pm$ 3.6  & $-1.07 \pm$ 0.07\\
\hline
0.7 $\leq z \leq$ 1.0\tablenotemark{b} & 32.8 $\pm$ 7.8 & $-0.92 \pm$ 0.11\\
1.0 $\leq z \leq$ 1.3\tablenotemark{b} & 65.8 $\pm$ 18.8 & $-0.61 \pm$ 0.14
\enddata
\tablenotetext{a}{\SFRD\ estimates assume a \citet{Salpeter55} IMF and
  are given in the unit of $M_{\odot} {\rm yr}^{-1} {\rm
    Mpc}^{-3}$. Assuming the \citet{Kroupa02} IMF would yield 0.66
  \SFRD\ given here.} 
\tablenotetext{b}{Based on FIDEL luminosity functions
  \citep{Magnelli09}.}
\label{table_evol}
\end{deluxetable}

\subsection{Extending the LF evolution constraint to $z \sim
  1.2$}\label{sec:results_LF_evol_to_FIDEL}
We augment our LF evolution constraint at low and intermediate
redshifts with the result of \citet{Magnelli09} at 35 $\micron$. This
is a useful combination because \citet{Magnelli09} used the deep (50
$\mu Jy$ and 3 mJy at 24 and 70 $\micron$, respectively), narrow-field
(0.5 sq. deg.), FIDEL survey. The FIDEL sample has $\sim 7000$ and
$\sim 600$ star-forming galaxies detected at 24 $\micron$ and 70
$\micron$, respectively, with known redshifts. AGN were excluded on
the basis of deep (200 ks - 2 Ms) X-ray observations. Additionally,
\citet{Magnelli09} employ a stacking analysis to provide further
constraints for the LFs. Their approach provides one of the best
evolution constraints for the LFs at higher redshifts to date but will
be affected by cosmic variance at intermediate redshifts, where our
results should be more reliable. In addition, the rest frame 35
$\micron$ values can be corrected to 24 $\micron$ with minimal uncertainties
due to luminosity conversions.

We need to correct \citet{Magnelli09} for (1) bandpass differences;
and (2) k-correction differences (SED assumptions). To correct for
bandpass differences, we use the \citet{Rieke09} SED library to
calculate a luminosity-dependent conversion factor between the
\citet{Magnelli09} rest-frame 35 $\micron$, which was observed at
70 $\micron$ and k-corrected to $z = 1$ (the observed 70
$\micron$ band probes the 35 $\micron$ rest-frame SED at $z =
1$). This conversion factor varies slowly from 0.5 at $\LTIR \sim
10^{10} \Lsun$ to 0.3 at $\LTIR \sim 10^{12} \Lsun$. Next we compare
the consistency of the \citet{CharyElbaz01} SED used by
\citet{Magnelli09} and our \citet{Rieke09} SED. The best test for
differences lies in the L$_*(z = 0)$ anchor point, L$_*(z = 0,
35~\micron) = 10^{9.85}$ $\Lsun$, that \citet{Magnelli09} obtained by
converting the \citet{Shupe98} IRAS 25 $\micron$ L$_*$ to 35 
$\micron$ (MIPS 70 $\micron$ observed bandpass at $z=1$) using the
\citet{CharyElbaz01} SED. Converting this L$_*(z = 0, 35~\micron)$
back to our MIPS 24 $\micron$ rest-frame (which is about 5\% fainter
than the original value at 25 $\micron$) using the \citet{Rieke09} SED
would yield L$_*(z = 0,24~\micron)$ of $3.72 \times 10^9$
$h^{-2}\Lsun$, which is 20\% lower than the original \citet{Shupe98}
value tabulated in Table \ref{table_comparellf}. In other words, a
20\% discrepancy resulted from subjecting a luminosity through this
conversion exercise and we must apply a $+20$\% correction factor to
the FIDEL data points to correct for SED differences.
 
For the combined fit, we omitted two data points. First, we dropped
the lowest redshift \citet{Magnelli09} bin ($0.4 \leq z \leq 0.7$), which
overlaps with our highest redshift bin, because this redshift bin will
be more strongly affected by cosmic variance since our highest 
redshift bin encompasses about an order of magnitude more
volume. Second, we dropped our local point because it was estimated by
evolving the sample in the lowest redshift bin ($0.05 \leq z \leq 0.20$)
back to $z=0.0$ based on the higher redshift bins
(Section \ref{sec:IR_LF_evolparam}), so including this point would ``double
count'' the data. Instead we use the \citet{Shupe98} local point converted
to 24 $\micron$. Note, however, that our local L$_*$ estimate is
consistent with \citet{Shupe98}. The combined fit (shown in Figure
\ref{LF_Lstar_evol}) yields a slightly shallower evolution law than
that estimated from the \Bootes\ field 
alone (Section \ref{sec:results_LF_evol}), with $L_{*,combined} \propto
(1+z)^{3.4 \pm 0.2}$. 

\subsection{The Evolution of the Star Formation Rate
  Density}\label{sec:results_SFRD} 

Now that we have constructed the LF and determined the luminosity
evolution from $z = 0.0$ to $1.2$ we can provide a complete
picture of the \SFRD\ evolution from $z = 0.0$ to $z \sim 1.2$. We
compute SFRs based on \citet{Kroupa02} IMF used by
\citet{Rieke09}. This leads to estimates of \SFRD\ a factor of 0.66
lower than those assuming a \citet{Salpeter55} IMF with a single
power-law slope of $-1.35$ from 0.1 to 100 $M_{\odot}$.

Our estimates for \SFRD\ in each redshift bin to $z \sim 1.2$,
including both our results and the adjusted values from
\citet{Magnelli09}, are tabulated in Table \ref{table_evol}, and shown
in Figure \ref{SFR_Madauplot} along with \SFRD\ measurements from other
wavelengths from the UV to radio. Based on the fit of the combined data
out to $z \sim 1.2$, our \SFRD\ evolution law is given by
\begin{equation}{\rm log}(\SFRD) = (3.5 \pm 0.2){\rm log}(1+z) -
  1.96 \pm 0.08
\end{equation}

At $z = 0$ our fit yields a local \SFRD\ of $(1.09 \pm 0.21) \times
10^{-2} M_{\odot} {\rm yr}^{-1}{\rm Mpc}^{-3}$ for \citet{Kroupa02},
which corresponds to \SFRD\ of $(1.65 \pm 0.32) \times 10^{-2}
M_{\odot} {\rm yr}^{-1}{\rm  Mpc}^{-3}$ for a \citet{Salpeter55}
IMF. Our estimate agrees with previous works on the local \SFRD\ that
have properly accounted for the internal extinction of the
galaxy. Without such corrections the \SFRD\ will be underestimated by
50 $-$ 70 \% \citep{Gallego95, TresseMad98, Hanish06, Salim07}. The
agreement between \SFRD\ estimates corrected for internal extinction
and our \SFRD\ derived from 24 $\micron$ luminosity suggests that
$\sim$half of the local star formation is obscured and the MIR
provides a direct and reliable means for SFR measurement.

Figure \ref{SFR_Madauplot} illustrates the good agreement of the
normalization of our \SFRD\ estimates and those derived from the UV
observation corrected for the effects of intrinsic extinction of the
target galaxies (normalization corrections can be as large as 0.7-dex;
e.g., Figure 5 of \citet{Schimi05} illustrates the extent of the required
correction for the UV observations). \citet{Hopkins04} reports the fit
to a compilation of the \SFRD\ that has been corrected for extinction
of the target galaxies with an assumption of  a luminosity-dependent
obscuration. The resulting best fit is given by ${\rm log}(\SFRD) =
(3.29 \pm 0.26){\rm log}(1+z) - 1.98 \pm 0.04$ (assuming the
\citet{Kroupa02} IMF), which also agrees well with our evolution
determined from the combination of the NDWFS and the FIDEL
results. Again, this agreement confirms that star formation at
redshift $0.0 < z < 1.2$ occurs in obscured environments and can be
traced directly with 24 $\micron$ observations given a proper
classification and exclusion of AGNs. It should be noted that our 
result is consistent with findings of \citet{Takeuchi05}, which show 
an evolution of the fraction of obscured star formation from 50\% 
locally to $> 80$\% at $z \sim 1$.

\section{CONCLUSIONS}\label{sec:conclusions}
We study the evolution of 24 $\micron$-selected galaxies by
constructing their 24 $\micron$ (rest-frame) luminosity functions
(LF). Our sample contains 4,047 galaxies with spectroscopic redshifts
at $0.0 \leq z \leq 0.65$ in the \Bootes\ field of the NOAO Deep-Wide
Field Survey (NDWFS). The 24 $\micron$ data and spectroscopic
redshifts were obtained with {\it Spitzer} and with the AGN and
Galaxies Evolution Survey (AGES), respectively. Our sample is a unique
combination of wide field, spectroscopic redshifts, and 24
$\micron$ imaging that is ideal for intermediate redshifts where MIR
galaxy evolution has not been well-studied. The large area (9
sq. deg.) of the \Bootes\ field helps mitigate cosmic variance,
which otherwise can be a serious issue at low and  intermediate
redshifts. Theoretically, the cosmic variance in our study is less
than 10\% at $z > 0.2$. Spectroscopic redshifts reduce 
cross-talk between redshift bins and faint/bright end slope biases and
provide an accurate number density contribution for each object. 

Galaxies exhibiting AGN activities characterized by X-ray emission
and mid-IR power-law are excluded from our star-forming sample. We
find that the optical line diagnostics (the BPT method) are
not suitable for identifying AGN dominant at mid-IR wavelengths
because the mid-IR emission from optically-selected AGNs is often
dominated by star formation. We identified 288 objects with 24
$\micron$ emission likely to be dominated by AGN. Excluding them
leaves a sample of 3,759 star-forming galaxies.

We derive the 24 $\micron$ luminosity using calibrations derived
from the most recent SED library of the mid-IR spectra from {\it
  Spitzer} \citep{Rieke09}. Our sample is comprised mainly of normal
star-forming galaxies (66\% have $\LTIR < 10^{11}$ \Lsun) and LIRGs
(33\% have $10^{11}$ \Lsun $< \LTIR < 10^{12}$ \Lsun). Only
1\% are ULIRGs ($\LTIR > 10^{12}$ \Lsun) and we found that most
of the ULIRGs contain AGN. 

We construct the local 24 $\micron$ luminosity function (LLF) as a
template to study evolution of the LF. Our LLF was constructed at
$0.05 \leq z  \leq 0.25$ and evolved back to $z = 0$ simultaneously
with the fit to constrain the global evolution of our sample. The LLF
template for our star-forming sample is given by a double power-law
with faint-end slope, $\alpha = 0.37 \pm 0.04$, bright-end slope,
$\beta = 2.36 \pm 0.41$, characteristic luminosity, L$_*(z=0) = (4.27
\pm 0.71) \times 10^9 \Lsun$, and density normalization, $C = (1.2 \pm
0.8) \times 10^{-3}{\rm Mpc}^{-3}$. Our major results are as follows.

1. The evolution of the LF at $z \leq 0.65$ can be represented by pure
luminosity evolution with the characteristic 24 $\micron$ luminosity
of star-forming galaxies evolving as L$_*(24~\micron) \propto
(1+z)^{3.8 \pm 0.3}$. We demonstrate by the construction of the
bivariate I-band and 24 $\micron$ luminosity function that the
fraction of missing 24 $\micron$-luminous, optically-faint galaxies
due to our I-band magnitude limit for spectroscopic targeting is very
small.

2. We extend the constraint on the evolution of L$_*$ to $z \sim 1.2$
by combining our results with the higher redshift results from
\citet{Magnelli09} based on the FIDEL survey. The combined sample
gives a slightly shallower luminosity evolution of L$_*(24~\micron)
\propto (1+z)^{3.4 \pm 0.2}$. The shallower evolution law derived by
including the FIDEL sample at redshifts of $z = 0.85$ and $1.15$
suggests that the evolution in \SFRD\ may begin to slow compared to
lower redshifts.

3. The local star formation rate density based on our 24 $\micron$
data is \SFRD\ of $(1.09 \pm 0.21) \times 10^{-2} M_{\odot} {\rm
  yr}^{-1}{\rm Mpc}^{-3}$ assuming the \citet{Kroupa02} IMF, which
corresponds to \SFRD\ of $(1.65 \pm 0.32) \times 10^{-2} M_{\odot} {\rm
  yr}^{-1}{\rm Mpc}^{-3}$ for a \citet{Salpeter55} IMF. The combined
evolution constraint for the  LF at $z \leq 1.2$ indicates that
\SFRD\ evolves as $\SFRD \propto (1+z)^{3.5 \pm 0.2}$. Our
normalization and evolution measurements agree well with the $z
\lesssim 1$ extinction-corrected findings from other studies, which
confirms that most of star formation is obscured and that 24 $\micron$
serves as a direct and reliable \SFRD\ indicator in these cases. 

We thank Benjamin Weiner, Benjamin Magnelli, Christopher Willmer, and
Pablo P{\'e}rez-Gonz{\'a}lez for invaluable discussions. W.R. thanks 
Andrew Hopkins for supplying his \SFRD\ compilation and Xinyu Dai for
his inputs on the cosmic variance in the \Bootes\ field. This work was
supported by the contract 1255094 from JPL/Caltech to the University
of Arizona. W.R. was supported by {\it Spitzer} archival grant
JPL-1278815, NSF AST-0607541, and the Thai Government Scholarship.


\begin{thebibliography}{}

\bibitem[Alonso-Herrero et al.(2006)]{AH06} Alonso-Herrero, 
A., Rieke, G.~H., Rieke, M.~J., Colina, L., P{\'e}rez-Gonz{\'a}lez, P.~G., 
\& Ryder, S.~D.\ 2006, \apj, 650, 835 

\bibitem[Ashby et al.(2009)]{Ashby09} Ashby, M.~L.~N., et al.\ 
2009, \apj, 701, 428 

\bibitem[Baldwin et al.(1981)]{BPT81} Baldwin, J.~A., 
Phillips, M.~M., \& Terlevich, R.\ 1981, \pasp, 93, 5 

\bibitem[Barmby et al.(2006)]{Barmby06} Barmby, P., et al.\ 
2006, \apj, 642, 126 

\bibitem[Bavouzet et 
al.(2008)]{Bavouzet08} Bavouzet, N., Dole, H., Le Floc'h, E., Caputi,
  K.~I., Lagache, G., \& Kochanek, C.~S.\ 2008, \aap, 479, 83  

\bibitem[Bertin \& Arnouts(1996)]{Bertin96} Bertin, E., \& Arnouts, S.\ 1996, 
\aaps, 117, 393 

\bibitem[Blanton \& Roweis(2007)]{Blanton07} Blanton, M.~R., \& 
Roweis, S.\ 2007, \aj, 133, 734 

\bibitem[Brand et al.(2006)]{Brand06} Brand, K., et al.\ 2006, 
\apj, 641, 140 

\bibitem[Brand et al.(2009)]{Brand09} Brand, K., et al.\ 2009, 
\apj, 693, 340 

\bibitem[Brown et al.(2007)]{Brown07} Brown, M.~J.~I., Dey, A., 
Jannuzi, B.~T., Brand, K., Benson, A.~J., Brodwin, M., Croton, D.~J., 
\& Eisenhardt, P.~R.\ 2007, \apj, 654, 858 

\bibitem[Buat et al.(1999)]{Buat99} Buat, V., Donas, J., 
Milliard, B., \& Xu, C.\ 1999, \aap, 352, 371 

\bibitem[Buat et al.(2007)]{Buat07} Buat, V., et al.\ 2007, 
\apjs, 173, 404 

\bibitem[Cardamone et al.(2008)]{Cardamone08} Cardamone, C.~N., et 
al.\ 2008, \apj, 680, 130 

\bibitem[Caputi et al.(2007)]{Caputi07} Caputi, K.~I., et al.\ 
2007, \apj, 660, 97 

\bibitem[Chary \& Elbaz(2001)]{CharyElbaz01} Chary, R., \& Elbaz, 
D.\ 2001, \apj, 556, 562 

\bibitem[Chary(2007)]{Chary07} Chary, R.-R.\ 2007, Deepest 
Astronomical Surveys, 380, 375 

\bibitem[Dai et al.(2009)]{Dai09} Dai, X., et al.\ 2009, 
\apj, 697, 506 

\bibitem[{{Dale} {et~al.}(2001){Dale}, {Helou}, {Contursi}, {Silbermann}, \&
  {Kolhatkar}}]{Dale01}
{Dale}, D.~A., {Helou}, G., {Contursi}, A., {Silbermann}, N.~A., \&
  {Kolhatkar}, S. 2001, \apj, 549, 215

\bibitem[{{Dale} \& {Helou}(2002)}]{DaleHelou02}
{Dale}, D.~A. \& {Helou}, G. 2002, \apj, 576, 159

\bibitem[Dale et al.(2007)]{Dale07} Dale, D.~A., et al.\ 2007, 
\apj, 655, 863 

\bibitem[Donley et al.(2007)]{Donley07} Donley, J.~L., Rieke, 
G.~H., P{\'e}rez-Gonz{\'a}lez, P.~G., Rigby, J.~R., 
\& Alonso-Herrero, A.\ 2007, \apj, 660, 167 

\bibitem[Donley et al.(2008)]{Donley08} Donley, J.~L., Rieke, 
G.~H., P{\'e}rez-Gonz{\'a}lez, P.~G., \& Barro, G.\ 2008, \apj, 687, 111 

\bibitem[Dole et al.(2004)]{Dole04} Dole, H., et al.\ 2004, 
\apjs, 154, 87 

\bibitem[Eisenhardt et al.(2004)]{Eisenhardt04} Eisenhardt, P.~R., 
et al.\ 2004, \apjs, 154, 48 

\bibitem[Elbaz et al.(1999)]{Elbaz99} Elbaz, D., et al.\ 1999, 
\aap, 351, L37

\bibitem[Elbaz et al.(2007)]{Elbaz07} Elbaz, D., et al.\ 2007, 
\aap, 468, 33 

\bibitem[Elvis et al.(1994)]{Elvis94} Elvis, M., et al.\ 1994, 
\apjs, 95, 1 

\bibitem[Fabricant et al.(2005)]{Fab05} Fabricant, D. et al.\ 2005,
\pasp, 117, 1411

\bibitem[Farrah et al.(2008)]{Farrah08} Farrah, D., et al.\ 
2008, \apj, 677, 957 

\bibitem[Franceschini et al.(2001)]{Frances01} Franceschini, A., 
Aussel, H., Cesarsky, C.~J., Elbaz, D., \& Fadda, D.\ 2001, \aap, 378, 1 

\bibitem[Gallego et al.(1995)]{Gallego95} Gallego, J., Zamorano, 
J., Aragon-Salamanca, A., \& Rego, M.\ 1995, \apjl, 455, L1 

\bibitem[Gordon et al.(2005)]{Gordon05} Gordon, K., et al.\ 2005, \pasp,
117, 503

\bibitem[Gorjian et al.(2008)]{Gorjian08} Gorjian, V., et al.\ 
2008, \apj, 679, 1040 

\bibitem[Hacking \& Houck(1987)]{HackingHouck87} Hacking, P., \&
  Houck, J.~R.\ 1987, \apjs, 63, 311

\bibitem[Hacking et al.(1987)]{Hacking87} Hacking, P., Houck, 
J.~R., \& Condon, J.~J.\ 1987, \apjl, 316, L15 

\bibitem[Hanish et al.(2006)]{Hanish06} Hanish, D.~J., et al.\ 
2006, \apj, 649, 150 

\bibitem[Hickox et al.(2009)]{Hickox09} Hickox, R.~C., et al.\ 
2009, \apj, 696, 891 

\bibitem[Hopkins(2004)]{Hopkins04} Hopkins, A.~M.\ 2004, \apj, 
615, 209 

\bibitem[Hopkins \& Beacom(2006)]{HopkinsBeacom06} Hopkins, A.~M., \& 
Beacom, J.~F.\ 2006, \apj, 651, 142 

\bibitem[Houck et al.(2005)]{Houck05} Houck, J.~R., et al.\ 
2005, \apjl, 622, L105 

\bibitem[{{Huchra} \& {Sargent}(1973)}]{Huchra73}
{Huchra}, J. \& {Sargent}, W.~L.~W. 1973, \apj, 186, 433

\bibitem[Ivezi{\'c} et al.(2002)]{Ivezic02} Ivezi{\'c}, {\v Z}., 
et al.\ 2002, \aj, 124, 2364 

\bibitem[Jannuzi \& Dey(1999)]{Jannuz99} Jannuzi, B.~T., \& Dey, 
A.\ 1999, Photometric Redshifts and the Detection of High Redshift 
Galaxies, 191, 111 

\bibitem[Kauffmann et al.(2003)]{Kauffmann03} Kauffmann, G., et 
al.\ 2003, \mnras, 346, 1055 

\bibitem[{{Kennicutt}(1998)}]{Kennicutt98}
{Kennicutt}, R.~C. 1998, \araa, 36, 189

\bibitem[Kewley et al.(2001)]{Kewley01} Kewley, L.~J., Dopita, 
M.~A., Sutherland, R.~S., Heisler, C.~A., \& Trevena, J.\ 2001, \apj, 556, 
121 

\bibitem[Kewley et al.(2006)]{Kewley07} Kewley, L.~J., Groves, 
B., Kauffmann, G., \& Heckman, T.\ 2006, \mnras, 372, 961 

\bibitem[Kroupa(2002)]{Kroupa02} Kroupa, P.\ 2002, Science, 295, 
82 

\bibitem[Lacy et al.(2004)]{Lacy04} Lacy, M., et al.\ 2004, 
\apjs, 154, 166 

\bibitem[Lagache et al.(2003)]{Lagache03} Lagache, G., Dole, H., 
\& Puget, J.-L.\ 2003, \mnras, 338, 555 

\bibitem[Lagache et al.(2004)]{Lagache04} Lagache, G., et al.\ 
2004, \apjs, 154, 112 

\bibitem[Le Floc'h et al.(2005)]{LeFloch05} Le Floc'h, E.,
et al.\ 2005, \apj, 632, 169

\bibitem[LeFloc'h et al.(2009)]{LeFloch09} LeFloc'h, E., et al.\ 
2009, \apj, 703, 222 

\bibitem[Leitherer et al.(1999)]{Starburst99} Leitherer, C., et 
al.\ 1999, \apjs, 123, 3 

\bibitem[{{Lilly} {et~al.}(1996){Lilly}, {Le Fevre}, {Hammer}, \&
  {Crampton}}]{Lilly96}
{Lilly}, S.~J., {Le Fevre}, O., {Hammer}, F., \& {Crampton}, D. 1996, \apjl,
  460, L1

\bibitem[Lilly et al.(2007)]{Lilly07} Lilly, S.~J., et al.\ 
2007, \apjs, 172, 70 

\bibitem[Madau et al.(1996)]{Mad96} Madau, P., Ferguson, 
H.~C., Dickinson, M.~E., Giavalisco, M., Steidel, C.~C., \& Fruchter, A.\ 
1996, \mnras, 283, 1388 

\bibitem[Magnelli et al.(2009)]{Magnelli09} Magnelli, B., Elbaz, D.,
  Chary, R.~R., Dickinson, M., Le Borgne, D., Frayer, D.~T., \&
  Willmer, C.~N.~A.\ 2009, \aap, 496, 57   

\bibitem[Marcillac et al.(2006)]{Marcillac06} Marcillac, D., Elbaz,
  D., Chary, R.~R., Dickinson, M., Galliano, F., \& Morrison,
  G.\ 2006, \aap, 451, 57

\bibitem[Martin et al.(2003)]{Martin03} Martin, C., et al.\ 
2003, \procspie, 4854, 336 

\bibitem[Marleau et al.(2004)]{Marleau04} Marleau, F.~R., et al.\ 
2004, \apjs, 154, 66 

\bibitem[Marleau et al.(2007)]{Marleau07} Marleau, F.~R., Fadda, 
D., Appleton, P.~N., Noriega-Crespo, A., Im, M., \& Clancy, D.\ 2007, \apj, 
663, 218 

\bibitem[Marshall et al.(1983)]{Marshall83} Marshall, H.~L., 
Tananbaum, H., Avni, Y., \& Zamorani, G.\ 1983, \apj, 269, 35 

\bibitem[Murray et al.(2005)]{Murray05} Murray, S.~S., et al.\ 
2005, \apjs, 161, 1 

\bibitem[Neugebauer et al.(1979)]{Neugebauer79} Neugebauer, G., Oke, 
J.~B., Becklin, E.~E., \& Matthews, K.\ 1979, \apj, 230, 79 

\bibitem[Papovich et al.(2004)]{Papovich04} Papovich, C., et al.\ 
2004, \apjs, 154, 70 

\bibitem[Papovich et al.(2006)]{Papovich06} Papovich, C., et al.\ 
2006, \aj, 132, 231 

\bibitem[Papovich et al.(2007)]{Papovich07} Papovich, C., et al.\ 
2007, \apj, 668, 45 

\bibitem[{{Peebles}(1980)}]{peebles80}
{Peebles}, P. 1980, The Large-Scale Structure of the Universe (Princeton
  University Press)

\bibitem[P{\'e}rez-Gonz{\'a}lez et al.(2005)]{PPG05} 
P{\'e}rez-Gonz{\'a}lez, P.~G., et al.\ 2005, \apj, 630, 82

\bibitem[Pozzi et al.(2004)]{Pozzi04} Pozzi, F., et al.\ 2004, 
\apj, 609, 122 

\bibitem[Reddy et al.(2008)]{Reddy08} Reddy, N.~A., Steidel, 
C.~C., Pettini, M., Adelberger, K.~L., Shapley, A.~E., Erb, D.~K., 
\& Dickinson, M.\ 2008, \apjs, 175, 48 

\bibitem[Rieke et al.(2004)]{Rieke04} Rieke, G.~H., et al.\ 
2004, \apjs, 154, 25 

\bibitem[Rieke et al.(2009)]{Rieke09} Rieke, G.~H., 
Alonso-Herrero, A., Weiner, B.~J., P{\'e}rez-Gonz{\'a}lez, P.~G., Blaylock, 
M., Donley, J.~L., \& Marcillac, D.\ 2009, \apj, 692, 556 

\bibitem[Rieke \& Lebofsky(1981)]{RiekeLebof81} Rieke, G.~H., \& Lebofsky, 
M.~J.\ 1981, \apj, 250, 87 

\bibitem[Rieke 
\& Lebofsky(1986)]{RiekeLebof86} Rieke, G.~H., \& Lebofsky,
  M.~J.\ 1986, \apj, 304, 326 

\bibitem[Rigby et al.(2008)]{Rigby08} Rigby, J.~R., et al.\ 
2008, \apj, 675, 262 

\bibitem[Rowan-Robinson et al.(1997)]{RowRob97} Rowan-Robinson, 
M., et al.\ 1997, \mnras, 289, 490 

\bibitem[Rush et al.(1993)]{Rush93} Rush, B., Malkan, M.~A., 
\& Spinoglio, L.\ 1993, \apjs, 89, 1 

\bibitem[Sandage et al.(1979)]{STY79} Sandage, A., Tammann, 
G.~A., \& Yahil, A.\ 1979, \apj, 232, 352 

\bibitem[Sanders et al.(2003)]{Sanders03} Sanders, D.~B., 
Mazzarella, J.~M., Kim, D.-C., Surace, J.~A., 
\& Soifer, B.~T.\ 2003, \aj, 126, 1607 

\bibitem[Salim et al.(2007)]{Salim07} Salim, S., et al.\ 2007, 
\apjs, 173, 267 

\bibitem[Salpeter(1955)]{Salpeter55} Salpeter, E.~E.\ 1955, \apj, 
121, 161 

\bibitem[Saunders et al.(1990)]{Saunders90} Saunders, W., 
Rowan-Robinson, M., Lawrence, A., Efstathiou, G., Kaiser, N., Ellis, R.~S., 
\& Frenk, C.~S.\ 1990, \mnras, 242, 318 

\bibitem[Schechter(1976)]{Schechter76} Schechter, P.\ 1976, \apj, 
203, 297 

\bibitem[Schiminovich et al.(2005)]{Schimi05} Schiminovich, D., 
et al.\ 2005, \apjl, 619, L47

\bibitem[Scoville et al.(2007)]{Scoville07} Scoville, N., et al.\ 
2007, \apjs, 172, 1 

\bibitem[Schlegel et al.(1998)]{SFD98} Schlegel, D.~J., 
Finkbeiner, D.~P., \& Davis, M.\ 1998, \apj, 500, 525 

\bibitem[{{Schmidt}(1968)}]{Schmidt68}{Schmidt}, M. 1968, \apj, 151, 393

\bibitem[Schmitt et al.(2006)]{Schmitt06} Schmitt, H.~R., 
Calzetti, D., Armus, L., Giavalisco, M., Heckman, T.~M., Kennicutt, R.~C., 
Jr., Leitherer, C., \& Meurer, G.~R.\ 2006, \apj, 643, 173 

\bibitem[Seymour et al.(2008)]{Seymour08} Seymour, N., et al.\ 
2008, \apjl, 681, L1 

\bibitem[Shi et al.(2007)]{Shi07} Shi, Y., et al.\ 2007, 
\apj, 669, 841 

\bibitem[Shupe et al.(1998)]{Shupe98} Shupe, D.~L., Fang, F., 
Hacking, P.~B., \& Huchra, J.~P.\ 1998, \apj, 501, 597 

\bibitem[Smith et al.(2007)]{JDSmith07} Smith, J.~D.~T., et al.\ 
2007, \apj, 656, 770 

\bibitem[Smol{\v c}i{\'c} et al.(2009)]{Smolcic09} Smol{\v 
c}i{\'c}, V., et al.\ 2009, \apj, 690, 610 

\bibitem[Soifer et al.(1989)]{Soifer89} Soifer, B.~T., Boehmer, 
L., Neugebauer, G., \& Sanders, D.~B.\ 1989, \aj, 98, 766 

\bibitem[Soifer \& Neugebauer(1991)]{SoiferNeu91} Soifer, B.~T., \& 
Neugebauer, G.\ 1991, \aj, 101, 354 

\bibitem[Stern et al.(2005)]{Stern05} Stern, D., et al.\ 2005, 
\apj, 631, 163

\bibitem[{{Stetson}(1987)}]{Stetson87}{Stetson}, P.~B. 1987, \pasp,
  99, 191

\bibitem[Takeuchi et 
al.(2005)]{Takeuchi05} Takeuchi, T.~T., Buat, V., \& Burgarella, D.\ 2005, \aap, 440, L17 


\bibitem[Tresse \& Maddox(1998)]{TresseMad98} Tresse, L., \& Maddox,
  S.~J.\ 1998, \apj, 495, 691 

\bibitem[Willmer et al.(2006)]{Willmer06} Willmer, C.~N.~A., et 
al.\ 2006, \apj, 647, 853 

\bibitem[{{Xu} {et~al.}(1998){Xu}, {Hacking}, {Fang}, {Shupe}, {Lonsdale},
  {Lu}, {Helou}, {Stacey}, \& {Ashby}}]{Xu98}
{Xu}, C., {Hacking}, P.~B., {Fang}, F., {et~al.} 1998, \apj, 508, 576

\bibitem[{{Xu}(2000)}]{Xu2000}{Xu}, C. 2000, \apj, 541, 134

\bibitem[Yahil et al.(1991)]{Yahil91} Yahil, A., Strauss, 
M.~A., Davis, M., \& Huchra, J.~P.\ 1991, \apj, 372, 380 

\bibitem[Zehavi et al.(2005)]{Zehavi05} Zehavi, I., et al.\ 
2005, \apj, 630, 1 

\bibitem[Zhu et al.(2009)]{Zhu09} Zhu, G., Moustakas, J., 
\& Blanton, M.~R.\ 2009, \apj, 701, 86 

\end{thebibliography}
\end{document}